\documentclass[aoas]{imsart}

\RequirePackage{amsthm,amsmath,amsfonts,amssymb,bm,amsthm, mathtools}
\RequirePackage[noend]{algpseudocode}
\RequirePackage{algorithm}
\algnewcommand{\algorithmicand}{\textbf{ and }} %\And command 
\algnewcommand{\And}{\algorithmicand} 
\RequirePackage{caption,subcaption,float,multirow}
\RequirePackage[authoryear]{natbib}
\RequirePackage[colorlinks,citecolor=blue,urlcolor=blue]{hyperref}%% uncomment this for coloring bibliography citations and linked URLs
\RequirePackage{graphicx, xcolor}%% uncomment this for including figures

\startlocaldefs
\numberwithin{equation}{section}
\theoremstyle{plain}
\theoremstyle{remark}
\DeclareMathOperator*{\argmin}{arg\,min}
\endlocaldefs

\begin{document}

\begin{frontmatter}
\title{Functional concurrent regression with compositional covariates and its application to the time-varying effect of causes of death on human longevity 
}
%\title{A sample article title with some additional note\thanksref{T1}}
\runtitle{Functional regression with compositional covariates}
%\thankstext{T1}{A sample of additional note to the title.}

\begin{aug}
%%%%%%%%%%%%%%%%%%%%%%%%%%%%%%%%%%%%%%%%%%%%%%%
%% Only one address is permitted per author. %%
%% Only division, organization and e-mail is %%
%% included in the address.                  %%
%% Additional information can be included in %%
\author[A]{\fnms{Emanuele Giovanni} \snm{Depaoli}\ead[label=e2,mark]{depaoli@stat.unipd.it}}
\author[B]{\fnms{Marco} \snm{Stefanucci}\ead[label=e1]{marco.stefanucci@uniroma2.it}},
\and
\author[A]{\fnms{Stefano} \snm{Mazzuco}\ead[label=e3,mark]{stefano.mazzuco@unipd.it}}
%%%%%%%%%%%%%%%%%%%%%%%%%%%%%%%%%%%%%%%%%%%%%%
%% Addresses                                %%
%%%%%%%%%%%%%%%%%%%%%%%%%%%%%%%%%%%%%%%%%%%%%% 
\address[A]{Department of Statistical Sciences,
University of Padova,
\printead{e2,e3}}

\address[B]{Department of Economics and Finance,
University of Rome Tor Vergata,
\printead{e1}}

\end{aug}

\begin{abstract}

\noindent
Multivariate functional data that are cross-sectionally compositional data are attracting increasing interest in the statistical modeling literature, a major example being trajectories over time of compositions derived from cause-specific mortality rates. In this work, we develop a novel functional concurrent regression model in which independent variables are functional compositions. This allows us to investigate the relationship over time between life expectancy at birth and compositions derived from cause-specific mortality rates of four distinct age classes, namely 0--4, 5--39, 40--64 and 65+ in 25 countries. A penalized approach is developed to estimate the regression coefficients and select the relevant variables. Then an efficient computational strategy based on an augmented Lagrangian algorithm is derived to solve the resulting optimization problem. The good performances of the model in predicting the response function and estimating the unknown functional coefficients are shown in a simulation study. The results on real data confirm the important role of neoplasms and cardiovascular diseases in determining life expectancy emerged in other studies and reveal several other contributions not yet observed.

\end{abstract}


\end{frontmatter}
%%%%%%%%%%%%%%%%%%%%%%%%%%%%%%%%%%%%%%%%%%%%%%
%% Please use \tableofcontents for articles %%
%% with 50 pages and more                   %%
%%%%%%%%%%%%%%%%%%%%%%%%%%%%%%%%%%%%%%%%%%%%%%
%\tableofcontents

%%%%%%%%%%%%%%%%%%%%%%%%%%%%%%%%%%%%%%%%%%%%%%
%%%% Main text entry area:

%%%%%%%%%%%%%%%%%%%%%%%%%%%%%%%%%%%%%%%%%%%%%%
%% INTRODUCTION
%%%%%%%%%%%%%%%%%%%%%%%%%%%%%%%%%%%%%%%%%%%%%%

\section{Introduction}
\label{sec:intro}
There is still a considerable heterogeneity across countries (even if we focus on high-income countries only) in terms of longevity, and the variability of the time pattern with which the recent mortality levels have been reached is even more heterogeneous. Several studies have investigated on these time patterns \citep[see, for instance,][]{canudas-romo2010}, but recently some are trying to analyze the role of causes of death in determining them. For example, \cite{Bergeron-Bouchere002414} try to determine which causes of death are associated with longevity extension. \cite{Woolf2019} attribute the recent stagnation of life expectancy in the USA to increasing midlife mortality caused by drug overdoses, alcohol abuses, suicides and some organ diseases. \cite{Mehta6998} contest these findings, arguing that cardiovascular diseases are mainly responsible for such stagnation. The idea of associating life expectancy (or other summary measures of mortality rates) with causes of death is not new: many investigators have employed a decomposition method (see, for instance, \cite{canudas-romo-vaupel2003}). However, most of these studies are limited to a single country (see \cite{Jasilonis2023,Mehta6998}) or to a single age group (see \cite{remund2018cause}). Others (see \cite{canudas2020reflection}) collapse time dimension into a single indicator, thus not considering the evolution of causes of death over the last decades. Recently \cite{Stefanucci2021} proposed a combination of Functional Data Analysis (FDA) and Compositional Data Analysis (CDA) to analyze the time pattern of causes of death, limiting to mortality at age 40--64. Although the study by \cite{Stefanucci2021} provides some useful insights on the evolution of cause-specific mortality, it remains of a descriptive nature and is limited to a specific age group, while it might be of interest to measure if and to what extent different compositions of causes of death are associated with the evolution of overall mortality in the latest years. Conducting such an analysis can prove highly beneficial in gaining valuable insights into the epidemiological experiences of different countries. Moreover, it allows for an indirect association with trends in risk factors, such as the prevalence of smoking.

We suggest that this can be performed by regressing the evolution of overall mortality (measured in terms of life expectancy at birth) with causes of death composition of mortality as defined by \cite{Stefanucci2021}. \cite{sun2020} have recently proposed a log-contrast regression model with functional compositional covariates but limited to the case of a scalar response variable. Although of great interest, their model is not specifically tailored to our purposes. Thus, we extend the previous work to cope with the functional essence of our response variable (life expectancy over time). Such an extension consists of a concurrent specification of the function-on-function linear regression model, with appropriate constraints due to the compositional nature of the covariates. Four age groups of causes of death are considered i.e., 0--4, 5--39, 40--64, and 65+, thus giving rise to four different compositions, each with many components -- not necessarily the same ones, as shown in Table \ref{tab:causes}. Since it is reasonable that only few of them are relevant to predict the outcome, the model specification assumes sparsity of the regression coefficients. In this way, variable selection is performed and interpretable results are obtained. An efficient computational strategy based on an augmented Lagrangian algorithm is also described to estimate the proposed model, and the performance of the method is illustrated through a simulation study.

The article proceeds as follows. In Section \ref{sec:data} we describe the analyzed data and formalize all the relevant quantities. In Section \ref{sec:methods} we introduce a novel concurrent functional regression model with compositional covariates and discuss its estimation. The results of a simulation study are presented in Section \ref{sec:sim} and the results on real data are extensively commented on Section \ref{sec:res}. Finally, Section \ref{sec:disc} concludes the article.

\begin{table}
\caption{Classifications of causes of death used and age groups for which they are considered.}
\centering
\resizebox{\columnwidth}{!}{ \begin{tabular}{p{12cm}|rrrr}
Classifications of causes of death &\multicolumn{4}{c}{Age classes}\\
\hline
    Congenital anomalies (CONG)                                   &0--4 & & & \\
    Infancy related causes, excluded congenital anomalies (INFA) &0--4 & & & \\
    Certain infectious and parasitic diseases (INFE)         &0--4 & 5--39 & 40--64 &65+ \\
    Neoplasms (NEOP)                                     &0--4 & 5--39 &40--64 &65+ \\
    Respiratory diseases (RESP)                                  &0--4 & 5--39 & 40--64 &65+ \\
    External causes of death (EXT)                              &0--4 & 5--39 & 40--64 &65+ \\
    Diseases of nervous system (NERV)                          &0--4 & 5--39 &40--64 &65+ \\
    Digestive system diseases (DIG)                       &  & 5--39    &40--64 &65+  \\
    Mental disorders (MENT)                                       &  & 5--39    &40--64 &65+  \\
    Endocrine, nutritional and metabolic diseases (END)          &   & 5--39   &40--64 &65+  \\
    Circulatory system diseases (CIRC)                        &   & 5--39   &40--64 &65+\\
    Diseases of urogenital system (UROG)                          &  &   &40--64 &65+  \\
    Lung cancer (LUNG)                                            &  &   &40--64 &65+  \\
    Diseases of skin, musculoskeletal system and connective tissue system (SKIN)     &   &  &40--64 &65+ \\
  \end{tabular}}
  \label{tab:causes}
\end{table}

% \begin{figure}
%     \centering
%     \includegraphics[scale=0.8]{Causes_example_PAA2022.pdf}
%     \caption{Compositions of causes of death in three selected countries for women, 1950--2018. Source: own elaboration of WHO mortality database}
%     \label{fig:curves_ex}
% \end{figure}

%%%%%%%%%%%%%%%%%%%%%%%%%%%%%%%%%%%%%%%%%%%%%%
%% DATA AND PROBLEM SETUP
%%%%%%%%%%%%%%%%%%%%%%%%%%%%%%%%%%%%%%%%%%%%%%

\section{Data and problem setup}
\label{sec:data}

\begin{table}
\caption{Considered countries.}
\centering
\begin{tabular}{l|l}
Area & Country \\
\hline 
North Eur.   &  Denmark, Finland, Norway, Sweden   \\
West Eur.  &  Austria, Belgium, Switzerland, France, Ireland, Netherlands, UK   \\
East Eur.   &  Hungary, Poland, Lithuania, Estonia, Latvia, Russia, Ukraine   \\
South Eur.   &  Italy, Spain \\
Extra Eur. & USA, Japan, New Zealand, Canada, Australia \\
\end{tabular}
\label{tab:countries}
\end{table}

For each cause $i$, age $x$ and calendar year $t$, we consider cause-specific mortality rates that can be written as
\begin{displaymath}
\prescript{i}{} {m}_{x}^{t}={m_x^t}\frac{\prescript{i}{}{D}_{x}^{t}}{D_x^t},
\end{displaymath}
where $\prescript{i}{}D_{x}^{t}$ is the number of deaths for cause $i$ at age $x$ and time $t$, $D_x^t$ is the number of deaths for all causes at age $x$ and time $t$ and $\prescript{i}{}m_{x}^{t}$ and $m_x^t$ are the corresponding rates. For a given age $x$, compositions of mortality rates can be regarded as compositions of $\prescript{i}{} {m}_{x}^{t}$ using ${m_x^t}$ as normalization constant. Otherwise, data with unit-sum constraints may be obtained from $\prescript{i}{}D_{x}^{t}$, using $\sum_{x,i}\prescript{i}{}D_{x}^{t}$ as the normalization constant. The latter approach was adopted by \citet{Oeppen2008CoherentFO} and \citet{kjaergaard2019forecasting} to model and forecast age-at-death distributions. In this way, the parts of the composition are related to different ages and the results could be difficult to interpret. Although this is not a problem for forecasting purposes, it is a major drawback for our perspective. The exact opposite of the previous approach is to study $\prescript{i}{} {m}_{x}^{t}$ directly, that is, different compositions for each age. This would result in many predictors, making estimation problematic, especially for limited sample sizes. Moreover, as before, interpreting the results could be challenging. For these reasons, we focus on four age classes: 0--4, 5--39, 40--64, 65+, giving rise to four different compositions. From a demographic point of view, they account for infant, premature, early-adult and senescent mortality causes of death patterns, respectively. The underlying idea is that not only does the cause-of-death composition change among age groups, but also its effect on life expectancy varies with age. Age stratification allows us to control for different age structures across countries and over time. Life expectancy is a summary indicator of overall mortality that is independent on age structure of population but the compositions of causes of death are potentially affected by age structure changes, since some causes might be negligible at very young ages and highly relevant for old ones (e.g. dementia) and others (e.g. congenital anomalies) may be the other way round. Therefore, by considering a distinct composition for each age group, we can take into account the changing significance of different causes of death according to age. Consequently, certain causes may become irrelevant for specific age classes. 

%Stratifying by age group also allow us to control for varying age structures of populations across countries and over time. Life expectancy is a summary indicator of overall mortality that is independent on age structure of population but the compositions of causes of death are potentially affected by varying age structure, since some causes might be negligible at very young ages and highly relevant for old ones (e.g. dementia) and others (e.g. congenital anomalies) may be the other way round. Therefore, by considering four distinct compositions for each age group, we can take into account the changing significance of different causes of death according to age. Consequently, certain causes may become irrelevant for specific age classes. 

Data on causes of death come from the WHO mortality database \citep{who} and from the Human Causes of Death database (HCD) \citep{hcod} which contain time series of age-specific and cause-specific deaths for several countries. A primary issue is that the International Classification of Diseases (ICD) has changed significantly over the years, which may bias results. Following \citet{canudas2020reflection} and \citet{Stefanucci2021}, we use broad categories of causes, which are minimally affected by the classification revisions. The categories considered are shown in Table \ref{tab:causes}: the number of causes is higher than in \cite{Stefanucci2021}, who limit their analysis to age group 40--64. Here, we also consider causes that are specific to infant ages (e.g., congenital anomalies) and senescent ones (e.g., mental disorders, including dementia and Alzheimer's disease). As can be seen in Table \ref{tab:causes}, only some of the 14 causes are included in the composition of a specific age group. For example, age 0--4 includes only 7 causes; the others are ignored as their role for that age group is negligible. On average, our classification accounts for $88\%$ of the total number of deaths for the age class 0--4, $92\%$ for the age class 5--39, and $98\%$ for the age classes 40--64 and 65+. Regarding the countries used in this work, after some preliminary analyses, we decided to limit the study to the $n=25$ nations reported in Table \ref{tab:countries}, with a population size exceeding one million and good data quality. In order to consider the same time window for each nation, we restrict the study to the years 1965--2012. Some years are still missing for a few countries, that is, 2005 for Australia, 1996–-1997 for Poland, and 2000 for the UK. This is not an issue, since our methodology also works for a non-equispaced time grid. Furthermore, a small number of zero counts is present in the age class 0--4 for external causes, neoplasms, infectious, respiratory and nervous diseases, as well as for mental and digestive diseases in the age class 5--39 and mental diseases in the other two age groups. Since the data have to be log-transformed, we replace them by the maximum rounding error of 0.5, which is a common practice in CDA \citep{aitchison2003manual}. Concerning life expectancy at birth, we use data from life tables from the Human Mortality Database (HMD) \citep{hmd}, which contains detailed, consistent, and high-quality data on human overall mortality, with no distinction among causes \citep{barbieri2015data}.

%%% METHODS-------------------------------------------------------------

\section{Methods}
\label{sec:methods}

The main objective is to analyze the time-varying effect of causes of death on human longevity, studying whether variations in the causes of death composition can be predictive of life expectancy at birth. Since life expectancies in a given year are calculated based on age-specific mortality rates for the same year, we assume a concurrent relationship between the response variable and the covariates. We formulate the statistical problem in a very general way, considering an arbitrary number $q$ of age classes and the possible inclusion of time-varying control variables (i.e., non-compositional covariates). Let $\bm{y}(t) = \left[ y_1(t),\ldots, y_n(t)\right]^\top \in \mathbb{R}^n$ be the response vector whose $i$-th component is the life expectancy at birth at time $t \in \mathcal{T}$ for the $i$-th country, with $i=1,\ldots,n$. Let ${\bm x}_{ij}(t) = \left[x_{ij1}(t),\ldots, x_{ijp_{j}}(t)\right]^{\top} \in \mathbb{S}^{p_j-1}$ be the composition of $p_j$ cause-specific mortality rates for the $i$-th nation and $j$-th age class at time $t$, with $j=1,\ldots,q$, and $\mathbb{S}^{p-1}=\left\{\left[x_1,\ldots, x_{p}\right]^\top \in \mathbb{R}^p, x_{k}>0, \sum_{{k}=1}^{p} x_{k} = 1 \right\}$ denoting the positive simplex lying in $\mathbb{R}^{p}$. Also, let ${\bm x}_{i}(t)=\left[\bm{x}_{i1}(t)^\top,\ldots, \bm{x}_{iq}(t)^\top\right]^\top \in \mathbb{R}^q$ be the vector containing all the $q$ compositions, with $p=\sum_{j=1}^q p_j$, and let ${\bm X}(t)=\left[\bm{x}_1(t),\ldots,\bm{x}_n(t)\right]^\top \in \mathbb{R}^{n \times p}$ be the matrix of functional predictors at time $t$. Finally, $\bm{Z}_c(t)\in \mathbb{R}^{n \times (p_c+1)}$ is the matrix of control variables at time $t$, where the first column is a vector of ones $\mathbf{1}_n$, to estimate the functional intercept. The observed life expectancies and compositions of mortality rates at each calendar year can be considered as discrete observations from $\bm{y}(t)$ and $\bm{X}(t)$, respectively. 

%In the next section, after a brief review of the linear log-contrast model in the scalar case, we introduce its extension to deal with functional covariates and response variable. 

\subsection{Linear log-contrast model}
\label{subsec:scalarmodel}
Since the pioneering work of \citet{aitchison1984log}, log-contrast models have been very popular for regression problems with compositional covariates. Suppose that we observe a response vector $\bm{y} = \left[y_1, \ldots, y_n\right]^\top \in \mathbb{R}^n$ and a design matrix $\bm{X}=\left[\bm{x}_1,\ldots, \bm{x}_n\right]^\top \in \mathbb{R}^{n\times p}$ with $\bm{x}_i=\left[x_{i1},\ldots, x_{ip}\right]^\top \in \mathbb{S}^{p-1}$, for $i=1\ldots,n$. Because of the unit-sum constraint, each row of the matrix $\bm{X}$ cannot vary freely and the classical regression model is subject to identification problems. A naive solution is to omit one of the parts of the composition, but the method is not invariant to the choice of the removed component and the resulting coefficients are difficult to interpret. The idea of \citet{aitchison1984log} is to perform an additive log-ratio transformation of the compositional data so that the transformed data admit the familiar Euclidean geometry in $\mathbb{R}^{p-1}$. For a given reference component $r\in\left\{1,\ldots, p\right\}$, let $\bm{Z}_r=\left[\bm{z}_1,\ldots, \bm{z}_n\right]^\top \in \mathbb{R}^{n \times (p-1)}$ be the associated design matrix, where the $j$-the element of $\bm{z}_i$ is given by $z_{ij}=\log\left(x_{ij}/x_{ir}\right)$, for $j=1,\ldots,r-1,r+1,\ldots,p$. The resulting linear log-contrast model is 
\begin{equation}
\label{eq:loglinref}
    \bm{y} = \mathbf{1}_n\beta_0+\bm{Z}_r\bm{\beta}_r+\bm{e},
\end{equation}
where $\beta_0$ is the intercept, $\bm{\beta}_r \in \mathbb{R}^{p-1}$ is the regression coefficient, and $\bm{e}\in \mathbb{R}^{n}$ is the error vector, independent from $\bm{Z}_r$ and distributed as $\mathcal{N}(0,\sigma^2)$. The log-contrast model can be written in the symmetric form
\begin{equation}
\label{eq:loglinsym}
    \bm{y} = \mathbf{1}_n\beta_0+\bm{Z}\bm{\beta}+\bm{e}, \quad \text{s.t. } \mathbf{1}_p^\top\bm{\beta}=0,
\end{equation}
where $\bm{Z}\in \mathbb{R}^{n \times p}$ is the matrix resulting from log-transforming each element of the matrix $\bm{X}$, $\beta_0$ and $\bm{e}$ are the same as in model (\ref{eq:loglinref}), and the regression coefficient $\bm{\beta}_r$ is the subvector obtained from $\bm{\beta}$ by removing the $r$-th component. The log-contrast model obeys a landmark concept in CDA, called subcompositional coherence \citep{aitchison2003manual}: if the $j$-th coefficient of $\bm{\beta}$ is zero, then the results are unchanged if the model is applied to the subcomposition without the $j$-th component. 

In the classical regression framework, the least squares estimation can be performed indifferently assuming the model (\ref{eq:loglinref}) or the constrained form in (\ref{eq:loglinsym}). However, in a high-dimensional setup where variable selection is required, the use of a Lasso penalization method \citep{tibshirani1996} determines the loss of equivalence between the symmetric and non-symmetric form. For example, consider the inclusion of a $L_1$ penalty term for model (\ref{eq:loglinref}), determining the optimization problem 
\begin{equation}
\label{eq:optref}
    \argmin_{\bm{\beta}_r, \beta_0}\dfrac{1}{2} \vert\vert \bm{y}-\mathbf{1}_n\beta_0-\bm{Z}_r\bm{\beta}_r \vert\vert^2_2+\lambda\vert\vert \bm{\beta}_r \vert\vert_1.
\end{equation}
The solution of problem (\ref{eq:optref}) is not invariant to the choice of the reference category $r$ and, in general, is different from that of the Lasso criteria related to the symmetric model (\ref{eq:loglinsym}), which determines the optimization problem 
\begin{equation}
\label{eq:optsym}
    \argmin_{\bm{\beta}, \beta_0}\dfrac{1}{2} \vert\vert \bm{y}-\mathbf{1}_n\beta_0-\bm{Z}\bm{\beta} \vert\vert^2_2+\lambda\vert\vert \bm{\beta}\vert\vert_1, \quad \text{s.t. } \mathbf{1}_p^\top\bm{\beta}=0.
\end{equation}
The latter is proposed and studied in the context of gut microbiome and metagenomic
data by \citet{lin2014variable}, who also provide theoretical guarantees for the resulting estimator. Moreover, the zero-sum constraint makes the model subcompositional coherent. 

\subsection{Sparse functional concurrent log-contrast regression}
\label{subsec:functModel}
Although in practice the functional compositional predictors and the response variable are observed at each calendar year, here we assume that $\bm{X}(t)$ and $\bm{y}(t)$ are observed for each $t\in\mathcal{T}$. Following the notation of Section \ref{subsec:scalarmodel} and Section \ref{sec:data}, let $\bm{Z}(t)\in \mathbb{R}^{n \times p}$ be the matrix resulting from log-transforming each element of the matrix $\bm{X}(t)$ at time $t$, and recall that $\bm{y}(t)\in\mathbb{R}^{n}$ is the functional response and $\bm{Z}_c(t)\in \mathbb{R}^{n \times (p_c+1)}$ is the functional matrix of control variables, including a vector of ones $\bm{1}_n$ as the first column. The matrix $\bm{Z}(t)$ contains $q$ compositions and thus we need to impose $q$ zero-sum constraints to achieve subcompositional coherence. Following \citet{lin2014variable} and \citet{sun2020}, we propose the functional concurrent log-contrast regression model 
\begin{equation} \label{eq:model_func}
     \bm{y}(t) = \bm{Z}_c(t)\bm{\beta}_c(t)+\bm{Z}(t)\bm{\beta}(t)+\bm{e}(t), \quad \text{s.t. } \bm{L}\bm{\beta}(t)=\bm{0}_q \quad \forall t \in \mathcal{T},
\end{equation}
where $\bm{\beta}(t)=\left[\bm{\beta}_1(t)^\top,\ldots,\bm{\beta}_q(t)^\top\right]^\top \in \mathbb{R}^p$ is the functional regression coefficient, with $\bm{\beta}_j(t)=[\beta_{j1}(t),\ldots, \beta_{jp_j}(t)]^\top \in \mathbb{R}^{p_j}$ for $j=1,\ldots,q$, $\bm{\beta}_c(t)\in\mathbb{R}^{p_c+1}$ is the functional regression coefficient related to the control variables, and $\bm{e}(t)\in\mathbb{R}^n$ is the vector of functional errors distributed as $\mathcal{N}(0,\sigma^2)$. The set of linear constraints is represented by the matrix 
\begin{equation*}
    \bm{L} = \begin{bmatrix}
    \mathbf{1}_{p_1} & \mathbf{0}_{p_1} & \cdots & \mathbf{0}_{p_1} \\
    \mathbf{0}_{p_2} & \mathbf{1}_{p_2} & \cdots & \mathbf{0}_{p_2} \\
    \vdots & \vdots & \ddots & \vdots \\
    \mathbf{0}_{p_q} & \mathbf{0}_{p_q} & \cdots & \mathbf{1}_{p_q} \\
  \end{bmatrix}^\top \in \mathbb{R}^{q\times p}. 
\end{equation*}

For our study, it is reasonable to assume that the effects of causes of death on life expectancy are smooth over years. To achieve smoothness, each coefficient curve is represented by a linear expansion of $k$ known basis functions, such that 
\begin{equation*}
    \bm{\beta}(t) = \bm{B}\bm{\Phi}(t), \quad \bm{\beta}_c(t) = \bm{B}_c\bm{\Phi}(t),
\end{equation*}
where $\bm{B}=\left[ \bm{b}_1, \ldots, \bm{b}_p\right]^\top \in \mathbb{R}^{p\times k}$ and $\bm{B}_c=\left[\bm{b}_0, \bm{b}_{c_1}, \ldots, \bm{b}_{c_p}\right]^\top \in \mathbb{R}^{(p_c+1)\times k}$ are the coefficient matrices, and $\bm{\Phi}(t)=\left[ \phi_1(t), \ldots, \phi_k(t)\right]^\top \in \mathbb{R}^k$ is the vector of basis functions. For simplicity and since it is usually sufficient in practice, here we assume the same number $k$ of basis functions for each predictor and control variable, obtained considering an equispaced grid of knots. Moreover, we assume that the elements of $\bm{\Phi}(t)$ are B-splines of order $d$ \citep{deboor1978}. A B-spline of order $d$ is a piecewise polynomial function of degree $d-1$ and is defined by a set of knots, which are the points where the functions meet. The choice is not restrictive, and other basis functions can be adopted: see \citet{ramsay2005} for a detailed discussion. The same consideration applies to the number of basis functions $k$, which can be assumed to be different for each coefficient curve.  

Another reasonable assumption is that some compositional components have no effect on life expectancy. To enable variable selection, we induce sparsity by using a $L_1$ penalization method. For model (\ref{eq:model_func}), the functional sparsity of the coefficient curves in $\bm{\beta}(t)$ translates into the row sparsity of the coefficient matrix $\bm{B}$. Many penalization methods have been proposed in Statistics and Machine Learning literature to induce sparsity, among which the Lasso \citep{tibshirani1996} is probably the most famous. The Group Lasso \citep{yuan2006glasso} is an extension which considers the concept of groups of coefficients and fits for the purpose here, since it allows the whole coefficient vectors $\bm{b}_{j}$, for $j=1,\ldots,p$, to be selected rather than their individual components. 

To formulate the optimization problem, the zero-sum constraints and the coefficient curves have to be expressed in terms of the elements of the matrices $\bm{B}$ and $\bm{B}_c$. For this purpose, it is convenient to express the problem in terms of $\bm{b}=\text{vec}(\bm{B}^\top) \in \mathbb{R}^{pk}$ and $\bm{b}_c=\text{vec}(\bm{B}_c^\top) \in \mathbb{R}^{(p_c+1)k}$. It can be easily seen that imposing $\mathbf{1}_{p_j}^\top\bm{b}_j(t)=0$, for $j=1,\ldots q$ and $\forall t \in \mathcal{T}$, is equivalent to imposing zero-sum constraints on the columns of the matrix $\bm{B}$, that is, $(\bm{L} \otimes \mathbf{I}_{k})\bm{b} = \tilde{\bm{L}}\bm{b}= \bm{0}_{qk}$ with $\tilde{\bm{L}} \in \mathbb{R}^{qk \times pk}$. Moreover, we have that
\begin{equation*}
    \bm{\beta}(t)=\left(\mathbf{I}_p \otimes \bm{\Phi}(t)^\top\right)\bm{b}=\tilde{\bm{\Phi}}(t)\bm{b},
\end{equation*}
with $\tilde{\bm{\Phi}}(t) \in \mathbb{R}^{p\times pk}$ and, similarly, $\bm{\beta}_c(t)=\tilde{\bm{\Phi}}_c(t)\bm{b}_c$, with $\tilde{\bm{\Phi}}_c(t) \in \mathbb{R}^{(p_c+1)\times (p_c+1)k}$. In accordance with the above considerations, we propose to estimate the parameters to solve the optimization problem
\begin{equation} \label{eq:opt_prob}
    \dfrac{1}{2}\argmin_{\bm{b}, \bm{b}_c}\int \bm{r}(t)^\top \bm{r}(t) dt + \lambda\sum_{j=1}^p \vert\vert \bm{b}_j \vert \vert_2, \quad 
    \text{s.t. } \tilde{\bm{L}}\bm{b} = \bm{0}_{qk},
\end{equation}
where $\bm{r}(t)=\bm{y}(t)-\bm{Z}_c(t)\tilde{\bm{\Phi}}(t)\bm{b}_c-\bm{Z}(t)\tilde{\bm{\Phi}}(t)\bm{b} \in \mathbb{R}^n$ and $\lambda$ is a tuning parameter that controls the strength of the group-Lasso penalization. The proposed estimator has the same desirable properties as its counterparts in the classical regression framework \citep{lin2014variable} and in the functional case with scalar response \citep{sun2020}. The zero-sum constraints for each composition guarantee that the estimator remains unchanged under the transformation $\bm{X}(t) \longmapsto \bm{S}\bm{X}(t)$, where $\bm{S}=\text{diag}(s_1,\ldots,s_n)$, with $s_i>0$ for $i=1,\ldots,n$. Furthermore, the constraints ensure that the proposed methodology is subcompositional coherent: if we knew that some coefficient curves of $\bm{\beta}(t)$ are zero and estimated the model using the compositions formed by excluding the parts associated with those curves, then the resulting estimator would be unchanged. Finally, a direct consequence of the symmetric formulation of the problem (\ref{eq:opt_prob}) is that the solution is invariant under any permutation of the components of each composition. 

\subsection{Computation}
We propose to solve the convex optimization problem (\ref{eq:opt_prob}) using an augmented Lagrangian algorithm \citep{bertsekas_constrained_1982}. For a detailed review of the method and its extensions with applications in Statistics and Machine Learning, see \citet{boyd2011admm}. The problem (\ref{eq:opt_prob}) can be rewritten as 
\begin{equation}
\label{eq:opt_prob2}
\begin{aligned}  
    \argmin_{\bm{b}, \bm{b}_c} & \frac{1}{2}\bm{b}^\top \bm{K} \bm{b}-\bm{b}^\top \bm{J} + \frac{1}{2}\bm{b}_c^\top \bm{M} \bm{b}_c-\bm{b}_c^\top \bm{P} + \bm{b}_c^\top \bm{Q} \bm{b} \\
    & + \lambda\sum_{j=1}^p \vert\vert \bm{b}_j \vert \vert_2, \quad 
    \text{s.t. } \tilde{\bm{L}}\bm{b} = \bm{0}_{qk},
\end{aligned}
\end{equation}
where the matrices containing functional inner products with weighting functions are denoted by $\bm{K} = \int\tilde{\bm{\Phi}}(t)^\top\bm{Z}(t)^\top \bm{Z}(t)\tilde{\bm{\Phi}}(t)dt \in \mathbb{R}^{pk \times pk}$, $\bm{J} = \int\tilde{\bm{\Phi}}(t)^\top\bm{Z}(t)^\top \bm{y}(t)dt \in  \mathbb{R}^{pk}$, $\bm{M} = \int \tilde{\bm{\Phi}}_c(t)^\top\bm{Z}_c(t)^\top \bm{Z}_c(t)\tilde{\bm{\Phi}}_c(t) dt \in \mathbb{R}^{(p_c+1) k \times (p_c+1) k}$, $\bm{P} = \int \tilde{\bm{\Phi}}_c(t)^\top\bm{Z}_c(t)^\top \bm{y}(t) dt \in \mathbb{R}^{(p_c+1) k}$ and $\bm{Q} = \int \tilde{\bm{\Phi}}_c(t)^\top \bm{Z}_c(t)^\top \bm{Z}(t)\tilde{\bm{\Phi}}(t) dt \in \mathbb{R}^{(p_c+1) k \times p_ck}$.

Since $\bm{b}_c$ is involved in neither the penalty term nor the constraint, the optimization problem can be restated as 
\begin{equation}
\label{eq:opt_prob3}
    \argmin_{\bm{b}} \frac{1}{2}\bm{b}^\top \tilde{\bm{K}} \bm{b}-\bm{b}^\top \tilde{\bm{J}} + \lambda\sum_{j=1}^p \vert\vert \bm{b}_j \vert \vert_2, \quad 
    \text{s.t. } \tilde{\bm{L}}\bm{b} = \bm{0}_{qk},
\end{equation}
where $\tilde{\bm{K}}=\bm{K}-\bm{Q}^\top \bm{M} ^{-1} \bm{Q}  \in \mathbb{R}^{pk \times pk}$ and $\tilde{\bm{J}} = \bm{J}-\bm{Q}^\top \bm{M}^{-1}\bm{P} \in \mathbb{R}^{pk}$. Then, once the solution $\widehat{\bm{b}}$ is obtained, the estimate of the coefficient associated with the control variables is $\widehat{\bm{b}}_c = \bm{M}^{-1}(\bm{P}-\bm{Q}\widehat{\bm{b}})$. 

The augmented Lagrangian associated with problem (\ref{eq:opt_prob3}) is 
\begin{equation*}
    L_\rho(\bm{b}, \bm{u}) = \dfrac{1}{2}\bm{b}^\top \tilde{\bm{K}} \bm{b}-\bm{b}^\top\tilde{\bm{J}}+\lambda \sum_{j=1}^p \vert\vert \bm{b}_j \vert\vert_2+\dfrac{\rho}{2}\vert\vert \tilde{\bm{L}}\bm{b}\vert\vert_2^2+\bm{u}^\top\tilde{\bm{L}}\bm{b},
\end{equation*}
where $\bm{u} \in \mathbb{R}^{qk}$ is the Lagrange multiplier and $\rho$ is the penalty parameter. The augmented Lagrangian method finds the solution of the original problem iterating between a minimization step and a dual ascent step. The procedure for a fixed $\lambda$ is summarized in Algorithm \ref{alg:alm}. We allow the penalty parameter $\rho$ to increase in each iteration if the error does not decrease sufficiently over the previous iteration. The adjustment scheme follows the guidelines in \citet[p. 123]{bertsekas_constrained_1982}. The first step of the algorithm updates
\begin{equation*}
     \bm{b}^k \gets \argmin_{\bm{b}} L_{\rho^{k-1}}(\bm{b}, \bm{u}^{k-1}),
\end{equation*}
and it is equivalent to solving a standard group-Lasso problem. In our implementation, we employ the Alternating Direction Method of Multipliers \citep{boyd2011admm}, but other routines can be used to solve the problem. When the model is fitted for a path of $\lambda$, the solutions $\widehat{\bm{u}}$ and $\widehat{\bm{b}}$ associated with the previous penalty term are used as a warm start for the subsequent iteration.

\begin{algorithm}
\caption{Augmented Lagrangian method to solve problem (\ref{eq:opt_prob3})}
\label{alg:alm}
\begin{algorithmic}
\Require $\bm{b}^0, \rho^0, \bm{u}^0, \epsilon, k_{\text{max}}$
\State $k \gets 1$ 
\State $\text{err}^0 \gets \max \tilde{\bm{L}}\bm{b}^0$ 
\While{$\text{err}^{k-1} > \epsilon \And k \le k_{\text{max}}$}
    \State $\bm{b}^k \gets \argmin_{\bm{b}} L_{\rho^{k-1}}(\bm{b}, \bm{u}^{k-1})$
    \State $\text{err}^k \gets \max \tilde{\bm{L}}\bm{b}^k$
    \If{$\text{err}^{k} > 0.25\text{err}^{k-1}$} 
        \State $\rho^k \gets 10 \rho^{k-1}$
    \Else
        \State $\rho^k \gets \rho^{k-1}$
        \State $\bm{u}^k \gets \bm{u}^{k-1}+\rho^k\tilde{L}\bm{b}^{k}$
    \EndIf
    \State $k \gets k+1$
\EndWhile
\end{algorithmic}
\end{algorithm}

As noted before, the functional compositional predictors and the response variable are observed at each calendar year and not continuously $\forall t \in \mathcal{T}$. Therefore, all the integrals involved in the optimization problem have to be computed from discrete-time observations. In our study, we employ the trapezoidal rule, which is equivalent to approximating the discrete-time data to continuous-time curves by means of linear interpolation.

\section{Simulations}\label{sec:sim}
We performed a simulation study in order to compare the performance of our proposal based on a constrained group Lasso (CGL) with two possible competitors. The first candidate is a baseline method, that is, a standard group Lasso in which the reference level $r$ is chosen randomly (BGL). The second competitor is based on a naive approach, which consists of estimating the log-contrast regression model with the Lasso penalty of \citet{lin2014variable} at each time $t \in \mathcal{T}$ and smoothing the resulting estimates. 

We generate the compositional data similarly to the previous works of \citet{lin2014variable}, \citet{shi_regression_2016}, \citet{sun2020}. The discrete-time grid is equispaced within the interval $\mathcal{T} = [0,1]$ and consists of $20$ time points $t_1, \ldots, t_{20}$. We consider scenarios with $q=4$ compositions, each with $p_j$ components, $j=1,\ldots,q$. To introduce dependence between the covariates, we use a compound symmetry correlation matrix $\bm{\Sigma}_X \in \mathbb{R}^{p_j \times p_j}$ with unit variances and correlations $\rho_X$. To account for time dependence, we consider a matrix $\bm{\Sigma}_T \in \mathbb{R}^{20 \times 20}$ with first-order autoregressive structure, unit variance and autoregressive parameter $\rho_T$. For each observation $i=1,\ldots,n$, the $j$-th composition over time is obtained by simulating $\bm{w}_{ij}=[\bm{w}_{ij}(t_1)^\top, \ldots, \bm{w}_{ij}(t_{20})^\top]^\top \sim \mathcal{N}(\bm{0}_{20p_j}, \sigma^2_X(\bm{\Sigma}_T\otimes\bm{\Sigma}_X))$ and then normalizing the counts as 
\begin{equation*}
    w_{ijl}(t_v) = \dfrac{\exp \left\{ w_{ijl}(t_v)\right\}}{\sum_{l=1}^{p_j} \exp \left\{ w_{ijl}(t_v)\right\}},
\end{equation*}
for $i=1,\ldots,n$, $l=1,\ldots,p_j$ and $v=1,\ldots,20$. The number of basis functions for cubic splines is set to $k=5$ and the number of components $p_j$ is the same across compositions and equal to $p/q$. Only 3 coefficients are non-null for each composition. The coefficient vectors are $\bm{b}_1=[1, -1, 0, 0, 0]^\top$, $\bm{b}_2=[ 0, 0, -0.5, 1, 0]^\top$, $\bm{b}_3=[-1, 1, 0.5, -1, 0]^\top$, $\bm{b}_{p_1+1}=[0.5, 0, 0, -0.5, 1]^\top$,  $\bm{b}_{p_1+2}=[ 0, 1, -1, 0, -1]^\top$, $\bm{b}_{p_1+3}= [-0.5, -1, 1, 0.5, 0]^\top$, $\bm{b}_{p_2+1}=[0.5, -1, -1, 1, 0]^\top$, $\bm{b}_{p_2+2}=[0, 1, 1, 0, 0]^\top$, $\bm{b}_{p_2+3}=[-0.5, 0, \linebreak[0] 0, -1, 0]^\top$, \linebreak $\bm{b}_{p_3+1}=[1, 0, 0.5, 0, -1]^\top$, $\bm{b}_{p_3+2}=[ 0, 0, -0.5, 0, 0]^\top$,  $\bm{b}_{p_3+3}=[-1, 0, 0, 0, 1]^\top$. We also consider scenarios with $p=40$ and $q=1$, with the same coefficients and the same degree of sparsity as for $p=40$ and $q=4$. For simplicity, we do not include either an intercept or other control variables. The response variables are generated from the model (\ref{eq:model_func}), with error terms distributed as $\mathcal{N}(0,\sigma^2)$, where $\sigma^2$ set to achieve specific signal-to-noise ratios (SNR). We simulated different settings $(n, p, q) = (50, 40, 1), (50, 40, 4), (50, 100, 4)$ and several combinations of parameters $\sigma_X^2 = 9$, $\rho_T = (0.2, 0.6)$, $\rho_X = (0.2, 0.6)$, $\text{SNR} = (2, 4)$. The tuning parameters $\lambda$ and $k$ are selected by ten-fold cross-validation and one-standard error rule \citep[p.~244]{hastie2009elements}

We use four different measures to compare our proposal with competitors. The prediction error is calculated using the average prediction mean square error $\sum_{v=1}^{20} \vert\vert \bm{y}(t_v)-\mathbf{1}_n^\top\widehat{\beta}_0(t_v)-\bm{Z}(t_v)\widehat{\bm{\beta}}(t_v) \vert\vert_2^2/(20n)$ computed from an independent test sample of size 1000. The estimation error is measured by $\sum_{j=1}^p \left(\int_{\mathcal{T}}\vert \widehat{\beta}_j(t)-\beta_j(t)\vert^2 dt\right)^{\frac{1}{2}}/p$. As variable selection measures, we use the false positive rate (FPR) and false negative rate (FNR), where positives and negatives refer to non-null and null coefficients, respectively. The naive method does not include a procedure for the selection of coefficient curves, but only a variable selection procedure at each time $t$, therefore, we select active predictors based on empirical evidence. Consequently, to have a fair comparison, we use the same criteria for all three methods. As in \citet{sun2020}, the estimated index set $\widehat{\mathcal{S}}$ of non-null coefficients is defined as
\begin{equation*}
    \widehat{\mathcal{S}} = \left\{j: \frac{\left(\int_{\mathcal{T}} \widehat{\beta}_j^2(t)dt\right)^{\frac{1}{2}}}{\sum_{j=1}^p\left(\int_{\mathcal{T}} \widehat{\beta}_j^2(t)dt\right)^{\frac{1}{2}}} \ge \frac{1}{p}, j=1,\ldots,p \right\}.
\end{equation*}

\begin{table}[]
\caption{Means and standard errors (in parentheses) of false positive and false negative rates
for the three methods with SNR = 2, based on 100 simulations.}
\resizebox{\columnwidth}{!}{
\begin{tabular}{lllllllllllll}
\hline
\multicolumn{5}{l}{Configuration}                             &  & \multicolumn{3}{l}{FPR($\%$)} &  & \multicolumn{3}{l}{FNR($\%$)} \\ \cline{1-5} \cline{7-9} \cline{11-13} 
$\rho_X$ & $\rho_T$ & $n$ & $p$ & $q$  &    & CGL     & BGL    & Naive&  & CGL     & BGL    & Naive    \\ \hline
$ 0.2 $ & $ 0.2 $ & $ 50 $ & $ 40 $ & $ 1 $ &  & $ 0.04 $ $ (0.04) $ & $ 0.39 $ $ (0.11) $ & $ 1.54 $ $ (0.20) $ &  & $ 3.58 $ $ (0.41) $ & $ 3.67 $ $ (0.42) $ & $ 10.42 $ $ (0.52) $ \\ 
  &   & 50 & 40 & 4 &  & $ 0.00 $ $ (0.00) $ & $ 0.36 $ $ (0.11) $ & $ 1.32 $ $ (0.23) $ &  & $ 2.67 $ $ (0.39) $ & $ 3.42 $ $ (0.41) $ & $ 10.67 $ $ (0.52) $ \\ 
  &   & 50 & 100 & 4 &  & $ 4.06 $ $ (0.22) $ & $ 7.18 $ $ (0.22) $ & $ 8.84 $ $ (0.30) $ &  & $ 0.25 $ $ (0.14) $ & $ 0.42 $ $ (0.18) $ & $ 8.83 $ $ (0.54) $ \\ 
$ 0.2 $ & $ 0.6 $ & $ 50 $ & $ 40 $ & $ 1 $ &  & $ 0.14 $ $ (0.07) $ & $ 0.64 $ $ (0.16) $ & $ 1.68 $ $ (0.25) $ &  & $ 3.50 $ $ (0.41) $ & $ 3.67 $ $ (0.42) $ & $ 10.67 $ $ (0.54) $ \\ 
  &   & 50 & 40 & 4 &  & $ 0.00 $ $ (0.00) $ & $ 0.79 $ $ (0.17) $ & $ 1.39 $ $ (0.23) $ &  & $ 3.42 $ $ (0.43) $ & $ 3.67 $ $ (0.43) $ & $ 11.08 $ $ (0.46) $ \\ 
  &   & 50 & 100 & 4 &  & $ 4.08 $ $ (0.20) $ & $ 7.28 $ $ (0.23) $ & $ 9.12 $ $ (0.33) $ &  & $ 0.75 $ $ (0.24) $ & $ 1.50 $ $ (0.32) $ & $ 8.58 $ $ (0.52) $ \\ 
$ 0.6 $ & $ 0.2 $ & $ 50 $ & $ 40 $ & $ 1 $ &  & $ 0.00 $ $ (0.00) $ & $ 0.32 $ $ (0.10) $ & $ 1.50 $ $ (0.20) $ &  & $ 3.92 $ $ (0.43) $ & $ 4.00 $ $ (0.45) $ & $ 10.00 $ $ (0.52) $ \\ 
  &   & 50 & 40 & 4 &  & $ 0.04 $ $ (0.04) $ & $ 0.64 $ $ (0.15) $ & $ 1.29 $ $ (0.22) $ &  & $ 2.83 $ $ (0.40) $ & $ 3.92 $ $ (0.45) $ & $ 10.58 $ $ (0.44) $ \\ 
  &   & 50 & 100 & 4 &  & $ 3.91 $ $ (0.19) $ & $ 6.98 $ $ (0.19) $ & $ 9.41 $ $ (0.34) $ &  & $ 0.92 $ $ (0.26) $ & $ 1.17 $ $ (0.29) $ & $ 9.00 $ $ (0.50) $ \\ 
$ 0.6 $ & $ 0.6 $ & $ 50 $ & $ 40 $ & $ 1 $ &  & $ 0.00 $ $ (0.00) $ & $ 0.71 $ $ (0.14) $ & $ 1.43 $ $ (0.23) $ &  & $ 4.33 $ $ (0.43) $ & $ 4.75 $ $ (0.43) $ & $ 10.67 $ $ (0.53) $ \\ 
  &   & 50 & 40 & 4 &  & $ 0.07 $ $ (0.05) $ & $ 0.93 $ $ (0.19) $ & $ 1.21 $ $ (0.20) $ &  & $ 3.08 $ $ (0.40) $ & $ 3.08 $ $ (0.40) $ & $ 10.33 $ $ (0.53) $ \\ 
  &   & 50 & 100 & 4 &  & $ 4.39 $ $ (0.20) $ & $ 7.57 $ $ (0.22) $ & $ 8.99 $ $ (0.32) $ &  & $ 0.92 $ $ (0.26) $ & $ 1.17 $ $ (0.29) $ & $ 9.92 $ $ (0.53) $ \\
\hline
\end{tabular}
\label{tab:snr2_1}
}
\end{table}

\begin{table}[]
\caption{Means and standard errors (in parentheses) of prediction and estimation errors
for the three methods with SNR = 2, based on 100 simulations. Estimation errors are multiplied by 100.}
\resizebox{\columnwidth}{!}{
\begin{tabular}{lllllllllllll}
\hline
\multicolumn{5}{l}{Configuration}                             &  & \multicolumn{3}{l}{Prediction error} &  & \multicolumn{3}{l}{Estimation error} \\ \cline{1-5} \cline{7-9} \cline{11-13} 
$\rho_X$ & $\rho_T$ & $n$ & $p$ & $q$  &    & CGL     & BGL    & Naive&  & CGL     & BGL    & Naive    \\ \hline
$ 0.2 $ & $ 0.2 $ & $ 50 $ & $ 40 $ & $ 1 $ &  & $ 8.45 $ $ (0.02) $ & $ 8.49 $ $ (0.02) $ & $ 13.92 $ $ (0.11) $ &  & $ 3.90 $ $ (0.04) $ & $ 3.97 $ $ (0.04) $ & $ 7.77 $ $ (0.04) $ \\ 
  &   & 50 & 40 & 4 &  & $ 8.22 $ $ (0.02) $ & $ 8.34 $ $ (0.03) $ & $ 12.55 $ $ (0.08) $ &  & $ 3.80 $ $ (0.04) $ & $ 4.15 $ $ (0.06) $ & $ 7.63 $ $ (0.05) $ \\ 
  &   & 50 & 100 & 4 &  & $ 8.35 $ $ (0.03) $ & $ 8.59 $ $ (0.04) $ & $ 14.64 $ $ (0.10) $ &  & $ 2.01 $ $ (0.02) $ & $ 2.25 $ $ (0.03) $ & $ 3.97 $ $ (0.02) $ \\ 
$ 0.2 $ & $ 0.6 $ & $ 50 $ & $ 40 $ & $ 1 $ &  & $ 8.46 $ $ (0.03) $ & $ 8.48 $ $ (0.03) $ & $ 14.04 $ $ (0.13) $ &  & $ 4.06 $ $ (0.05) $ & $ 4.13 $ $ (0.06) $ & $ 7.90 $ $ (0.06) $ \\ 
  &   & 50 & 40 & 4 &  & $ 8.35 $ $ (0.03) $ & $ 8.51 $ $ (0.03) $ & $ 12.63 $ $ (0.09) $ &  & $ 3.88 $ $ (0.04) $ & $ 4.25 $ $ (0.06) $ & $ 7.64 $ $ (0.06) $ \\ 
  &   & 50 & 100 & 4 &  & $ 8.35 $ $ (0.03) $ & $ 8.68 $ $ (0.04) $ & $ 14.64 $ $ (0.09) $ &  & $ 2.03 $ $ (0.02) $ & $ 2.31 $ $ (0.03) $ & $ 3.98 $ $ (0.03) $ \\ 
$ 0.6 $ & $ 0.2 $ & $ 50 $ & $ 40 $ & $ 1 $ &  & $ 4.18 $ $ (0.01) $ & $ 4.21 $ $ (0.01) $ & $ 7.33 $ $ (0.07) $ &  & $ 3.91 $ $ (0.04) $ & $ 4.00 $ $ (0.04) $ & $ 7.87 $ $ (0.05) $ \\ 
  &   & 50 & 40 & 4 &  & $ 4.08 $ $ (0.01) $ & $ 4.14 $ $ (0.01) $ & $ 6.24 $ $ (0.04) $ &  & $ 3.83 $ $ (0.04) $ & $ 4.12 $ $ (0.05) $ & $ 7.56 $ $ (0.04) $ \\ 
  &   & 50 & 100 & 4 &  & $ 4.30 $ $ (0.02) $ & $ 4.43 $ $ (0.02) $ & $ 7.62 $ $ (0.06) $ &  & $ 1.96 $ $ (0.02) $ & $ 2.21 $ $ (0.02) $ & $ 3.99 $ $ (0.02) $ \\ 
$ 0.6 $ & $ 0.6 $ & $ 50 $ & $ 40 $ & $ 1 $ &  & $ 4.22 $ $ (0.01) $ & $ 4.24 $ $ (0.01) $ & $ 7.23 $ $ (0.08) $ &  & $ 3.93 $ $ (0.04) $ & $ 4.06 $ $ (0.05) $ & $ 7.86 $ $ (0.06) $ \\ 
  &   & 50 & 40 & 4 &  & $ 4.04 $ $ (0.01) $ & $ 4.11 $ $ (0.01) $ & $ 6.17 $ $ (0.03) $ &  & $ 3.86 $ $ (0.04) $ & $ 4.21 $ $ (0.06) $ & $ 7.61 $ $ (0.04) $ \\ 
  &   & 50 & 100 & 4 &  & $ 4.34 $ $ (0.01) $ & $ 4.48 $ $ (0.02) $ & $ 7.59 $ $ (0.06) $ &  & $ 2.01 $ $ (0.02) $ & $ 2.29 $ $ (0.02) $ & $ 3.97 $ $ (0.02) $ \\
\hline
\end{tabular}
\label{tab:snr2_2}
}
\end{table}

The means and standard errors of the performance measures for the scenario with SNR = 2 are reported in Tables \ref{tab:snr2_1} and \ref{tab:snr2_2}. From Table \ref{tab:snr2_1}, we can see that the proposed CGL has a similar variable selection performance compared to BGL when $n>p$, although the latter has the tendency to have higher false positive rates. This behavior is due to the automatic inclusion of the randomly chosen baseline for BGL and is, in fact, even more pronounced for $q=4$. The advantages of the proposed CGL can be appreciated for the scenarios with $p>n$, where it clearly outperforms competitors. As seen in Table \ref{tab:snr2_2}, the proposed CGL performs slightly better in terms of prediction and estimation error and, as before, the difference with the competitors is emphasized for $p>n$. Furthermore, increasing the correlation between the components leads to lower prediction errors, regardless of the method. This is because a small correlation determines few dominating components in each composition. As expected, the naive method has inferior performance in terms of all the measures in all the settings, since it is an unsophisticated approximation of the functional nature of the data. Another expected behavior can be seen from Tables \ref{tab:snr4_1} and \ref{tab:snr4_2}, which show that increasing the SNR leads to improved performance.

\begin{table}[]
\caption{Means and standard errors (in parentheses) of false positive and false negative rates
for the three methods with SNR = 4, based on 100 simulations.}
\resizebox{\columnwidth}{!}{
\begin{tabular}{lllllllllllll}
\hline
\multicolumn{5}{l}{Configuration}                             &  & \multicolumn{3}{l}{FPR($\%$)} &  & \multicolumn{3}{l}{FNR($\%$)} \\ \cline{1-5} \cline{7-9} \cline{11-13} 
$\rho_X$ & $\rho_T$ & $n$ & $p$ & $q$  &    & CGL     & BGL    & Naive&  & CGL     & BGL    & Naive    \\ \hline
$ 0.2 $ & $ 0.2 $ & $ 50 $ & $ 40 $ & $ 1 $ &  & $ 0.00 $ $ (0.00) $ & $ 0.00 $ $ (0.00) $ & $ 0.21 $ $ (0.09) $ &  & $ 1.75 $ $ (0.34) $ & $ 1.50 $ $ (0.32) $ & $ 7.17 $ $ (0.43) $ \\ 
  &   & 50 & 40 & 4 &  & $ 0.00 $ $ (0.00) $ & $ 0.00 $ $ (0.00) $ & $ 0.18 $ $ (0.08) $ &  & $ 1.33 $ $ (0.31) $ & $ 1.83 $ $ (0.35) $ & $ 7.83 $ $ (0.33) $ \\ 
  &   & 50 & 100 & 4 &  & $ 1.14 $ $ (0.11) $ & $ 3.48 $ $ (0.17) $ & $ 4.57 $ $ (0.23) $ &  & $ 0.00 $ $ (0.00) $ & $ 0.00 $ $ (0.00) $ & $ 5.08 $ $ (0.51) $ \\ 
$ 0.2 $ & $ 0.6 $ & $ 50 $ & $ 40 $ & $ 1 $ &  & $ 0.00 $ $ (0.00) $ & $ 0.00 $ $ (0.00) $ & $ 0.14 $ $ (0.07) $ &  & $ 2.50 $ $ (0.38) $ & $ 2.33 $ $ (0.38) $ & $ 6.67 $ $ (0.44) $ \\ 
  &   & 50 & 40 & 4 &  & $ 0.00 $ $ (0.00) $ & $ 0.04 $ $ (0.04) $ & $ 0.11 $ $ (0.06) $ &  & $ 1.33 $ $ (0.31) $ & $ 1.75 $ $ (0.34) $ & $ 7.25 $ $ (0.37) $ \\ 
  &   & 50 & 100 & 4 &  & $ 1.37 $ $ (0.13) $ & $ 3.93 $ $ (0.18) $ & $ 5.12 $ $ (0.22) $ &  & $ 0.17 $ $ (0.12) $ & $ 0.08 $ $ (0.08) $ & $ 6.25 $ $ (0.49) $ \\ 
$ 0.6 $ & $ 0.2 $ & $ 50 $ & $ 40 $ & $ 1 $ &  & $ 0.00 $ $ (0.00) $ & $ 0.07 $ $ (0.05) $ & $ 0.18 $ $ (0.08) $ &  & $ 2.58 $ $ (0.39) $ & $ 2.67 $ $ (0.39) $ & $ 7.50 $ $ (0.37) $ \\ 
  &   & 50 & 40 & 4 &  & $ 0.00 $ $ (0.00) $ & $ 0.00 $ $ (0.00) $ & $ 0.00 $ $ (0.00) $ &  & $ 1.58 $ $ (0.33) $ & $ 1.58 $ $ (0.33) $ & $ 8.33 $ $ (0.37) $ \\ 
  &   & 50 & 100 & 4 &  & $ 1.30 $ $ (0.13) $ & $ 3.53 $ $ (0.17) $ & $ 4.23 $ $ (0.19) $ &  & $ 0.00 $ $ (0.00) $ & $ 0.17 $ $ (0.12) $ & $ 5.17 $ $ (0.46) $ \\ 
$ 0.6 $ & $ 0.6 $ & $ 50 $ & $ 40 $ & $ 1 $ &  & $ 0.00 $ $ (0.00) $ & $ 0.00 $ $ (0.00) $ & $ 0.14 $ $ (0.07) $ &  & $ 2.67 $ $ (0.39) $ & $ 2.67 $ $ (0.39) $ & $ 7.08 $ $ (0.46) $ \\ 
  &   & 50 & 40 & 4 &  & $ 0.00 $ $ (0.00) $ & $ 0.04 $ $ (0.04) $ & $ 0.11 $ $ (0.06) $ &  & $ 1.75 $ $ (0.34) $ & $ 1.83 $ $ (0.35) $ & $ 7.67 $ $ (0.35) $ \\ 
  &   & 50 & 100 & 4 &  & $ 1.24 $ $ (0.13) $ & $ 3.98 $ $ (0.18) $ & $ 5.28 $ $ (0.23) $ &  & $ 0.17 $ $ (0.12) $ & $ 0.33 $ $ (0.16) $ & $ 5.67 $ $ (0.49) $ \\
\hline
\end{tabular}
\label{tab:snr4_1}
}
\end{table}

\begin{table}[]
\caption{Means and standard errors (in parentheses) of prediction and estimation errors
for the three methods with SNR = 4, based on 100 simulations. Estimation errors are multiplied by 100.}
\resizebox{\columnwidth}{!}{
\begin{tabular}{lllllllllllll}
\hline
\multicolumn{5}{l}{Configuration}                             &  & \multicolumn{3}{l}{Prediction error} &  & \multicolumn{3}{l}{Estimation error} \\ \cline{1-5} \cline{7-9} \cline{11-13} 
$\rho_X$ & $\rho_T$ & $n$ & $p$ & $q$  &    & CGL     & BGL    & Naive&  & CGL     & BGL    & Naive    \\ \hline
$ 0.2 $ & $ 0.2 $ & $ 50 $ & $ 40 $ & $ 1 $ &  & $ 4.29 $ $ (0.01) $ & $ 4.31 $ $ (0.01) $ & $ 7.71 $ $ (0.08) $ &  & $ 2.98 $ $ (0.03) $ & $ 3.05 $ $ (0.03) $ & $ 5.97 $ $ (0.04) $ \\ 
  &   & 50 & 40 & 4 &  & $ 4.30 $ $ (0.01) $ & $ 4.36 $ $ (0.02) $ & $ 6.78 $ $ (0.05) $ &  & $ 2.85 $ $ (0.03) $ & $ 3.08 $ $ (0.04) $ & $ 5.71 $ $ (0.05) $ \\ 
  &   & 50 & 100 & 4 &  & $ 4.31 $ $ (0.01) $ & $ 4.46 $ $ (0.02) $ & $ 8.75 $ $ (0.09) $ &  & $ 1.53 $ $ (0.02) $ & $ 1.71 $ $ (0.02) $ & $ 3.28 $ $ (0.02) $ \\ 
$ 0.2 $ & $ 0.6 $ & $ 50 $ & $ 40 $ & $ 1 $ &  & $ 4.30 $ $ (0.02) $ & $ 4.33 $ $ (0.02) $ & $ 7.90 $ $ (0.11) $ &  & $ 3.00 $ $ (0.03) $ & $ 3.09 $ $ (0.03) $ & $ 6.02 $ $ (0.05) $ \\ 
  &   & 50 & 40 & 4 &  & $ 4.28 $ $ (0.01) $ & $ 4.34 $ $ (0.02) $ & $ 6.71 $ $ (0.06) $ &  & $ 2.86 $ $ (0.03) $ & $ 3.09 $ $ (0.04) $ & $ 5.69 $ $ (0.05) $ \\ 
  &   & 50 & 100 & 4 &  & $ 4.46 $ $ (0.02) $ & $ 4.61 $ $ (0.02) $ & $ 8.90 $ $ (0.08) $ &  & $ 1.52 $ $ (0.02) $ & $ 1.72 $ $ (0.02) $ & $ 3.29 $ $ (0.02) $ \\ 
$ 0.6 $ & $ 0.2 $ & $ 50 $ & $ 40 $ & $ 1 $ &  & $ 2.21 $ $ (0.01) $ & $ 2.22 $ $ (0.01) $ & $ 4.25 $ $ (0.05) $ &  & $ 2.93 $ $ (0.03) $ & $ 3.02 $ $ (0.03) $ & $ 6.00 $ $ (0.04) $ \\ 
  &   & 50 & 40 & 4 &  & $ 2.17 $ $ (0.01) $ & $ 2.19 $ $ (0.01) $ & $ 3.49 $ $ (0.03) $ &  & $ 2.85 $ $ (0.03) $ & $ 3.04 $ $ (0.04) $ & $ 5.72 $ $ (0.04) $ \\ 
  &   & 50 & 100 & 4 &  & $ 2.22 $ $ (0.01) $ & $ 2.28 $ $ (0.01) $ & $ 4.57 $ $ (0.04) $ &  & $ 1.51 $ $ (0.02) $ & $ 1.68 $ $ (0.02) $ & $ 3.28 $ $ (0.02) $ \\ 
$ 0.6 $ & $ 0.6 $ & $ 50 $ & $ 40 $ & $ 1 $ &  & $ 2.18 $ $ (0.01) $ & $ 2.18 $ $ (0.01) $ & $ 4.15 $ $ (0.05) $ &  & $ 3.02 $ $ (0.04) $ & $ 3.07 $ $ (0.04) $ & $ 6.03 $ $ (0.04) $ \\ 
  &   & 50 & 40 & 4 &  & $ 2.11 $ $ (0.01) $ & $ 2.14 $ $ (0.01) $ & $ 3.36 $ $ (0.03) $ &  & $ 2.92 $ $ (0.03) $ & $ 3.17 $ $ (0.04) $ & $ 5.81 $ $ (0.04) $ \\ 
  &   & 50 & 100 & 4 &  & $ 2.24 $ $ (0.01) $ & $ 2.32 $ $ (0.01) $ & $ 4.61 $ $ (0.05) $ &  & $ 1.53 $ $ (0.02) $ & $ 1.72 $ $ (0.02) $ & $ 3.30 $ $ (0.02) $ \\
\hline
\end{tabular}
\label{tab:snr4_2}
}
\end{table}

%%% Results-----------------------------------------------------------

\section{Results}\label{sec:res}

The proposed functional concurrent regression model with compositional covariates is fitted separately for males and females, since the trajectories of their causes of death have profoundly different characteristics. We use cubic spline bases, and the penalty parameter $\lambda$ as well as the number of basis functions $k$ are selected through leave-one-out cross-validation, due to the limited sample size, and one-standard error rule. In this application, the only control variable is the intercept.

As a by-product of the regression model results, we can measure the relative importance of causes in the $j$-th age class by considering the relative squared $L_2$ norm of the group-specific coefficients between years $t$ and $t+1$

\begin{equation*}
     \label{eq:rel_magn}
     \left. \sum_{l=1}^{p_j} \int_{t}^{t+1}\vert\beta_{jl}(t)\vert^2 dt \middle / \sum_{j=1}^4\sum_{l=1}^{p_j} \int_{t}^{t+1}\vert\beta_{jl}(t)\vert^2dt \right..
\end{equation*}

\begin{figure}
    \centering
    \includegraphics[width=0.8\textwidth]{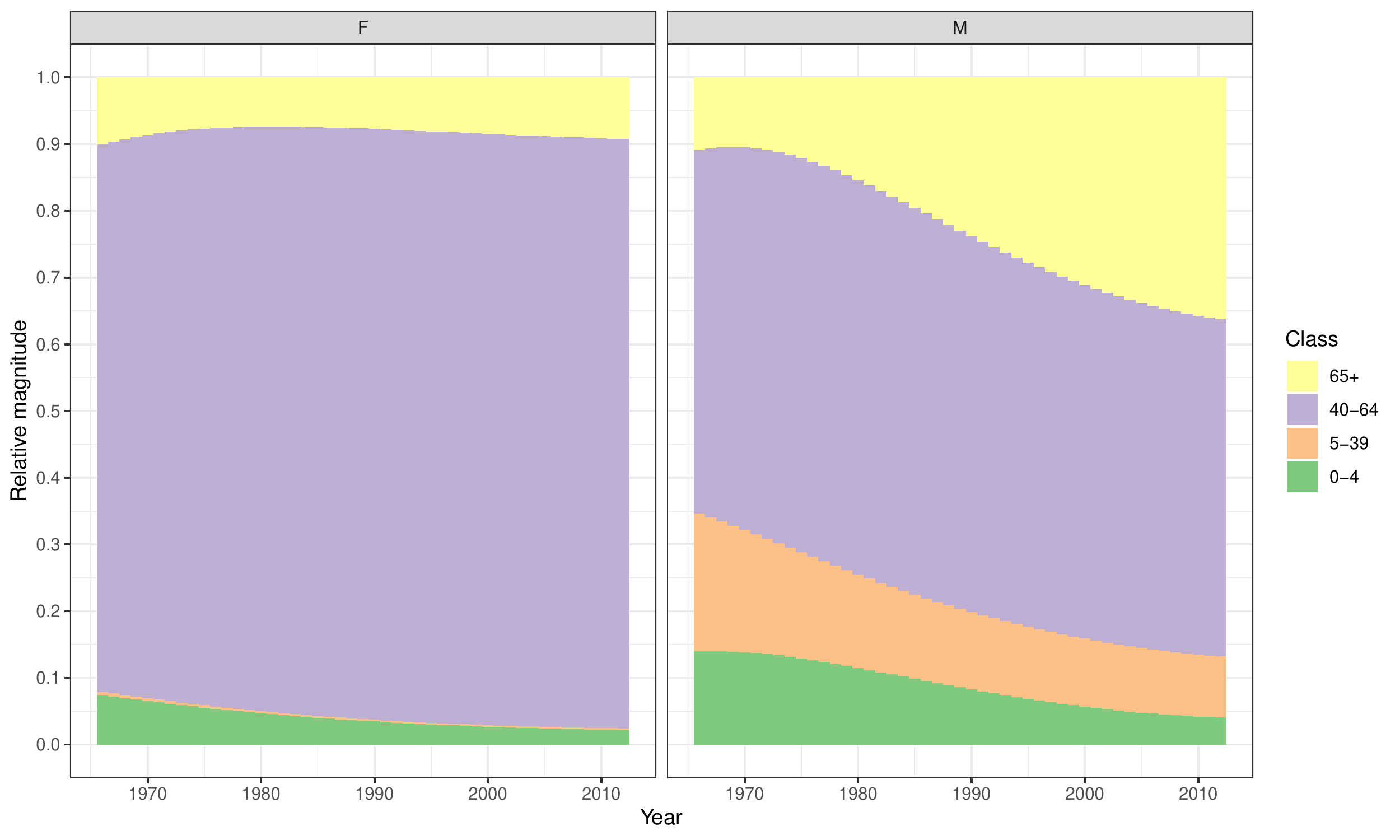}
    \caption{\textit{Relative magnitude of the age group--specific coefficients for females and males.}}
    \label{fig:rel_magn}
\end{figure}

The results are reported in Figure \ref{fig:rel_magn} and show that, for both men and women, the most important age class is 40--64. This can be attributed to the inclusion of countries from Eastern Europe, for which the compositional trajectories in the age group 40--64 are very different from the other high-longevity nations. The result is consistent with the demographic literature, in which traditional life expectancy decomposition methods are applied longitudinally for single countries. For example, \citet{mesle2004mortality} shows that in many former Soviet countries, decreases in life expectancy in the period 1965--2000 for males can be attributed to the rise in mortality at working ages. This is also in line with the substantial sex difference in the contribution of the age group 5--39. Another expected finding is the decline in importance for the age group 0--4, regardless of sex, which is associated with a progressive reduction in infant mortality. We also notice an increasing importance of age class 65+ for men. This can be explained by the faster progress of men in reducing heart disease-related mortality in recent decades, a pattern observed by \cite{feraldi-zarrulli2022}. 

\begin{figure}
\begin{subfigure}{0.45\textwidth}
  \centering
  \includegraphics[width=\linewidth]{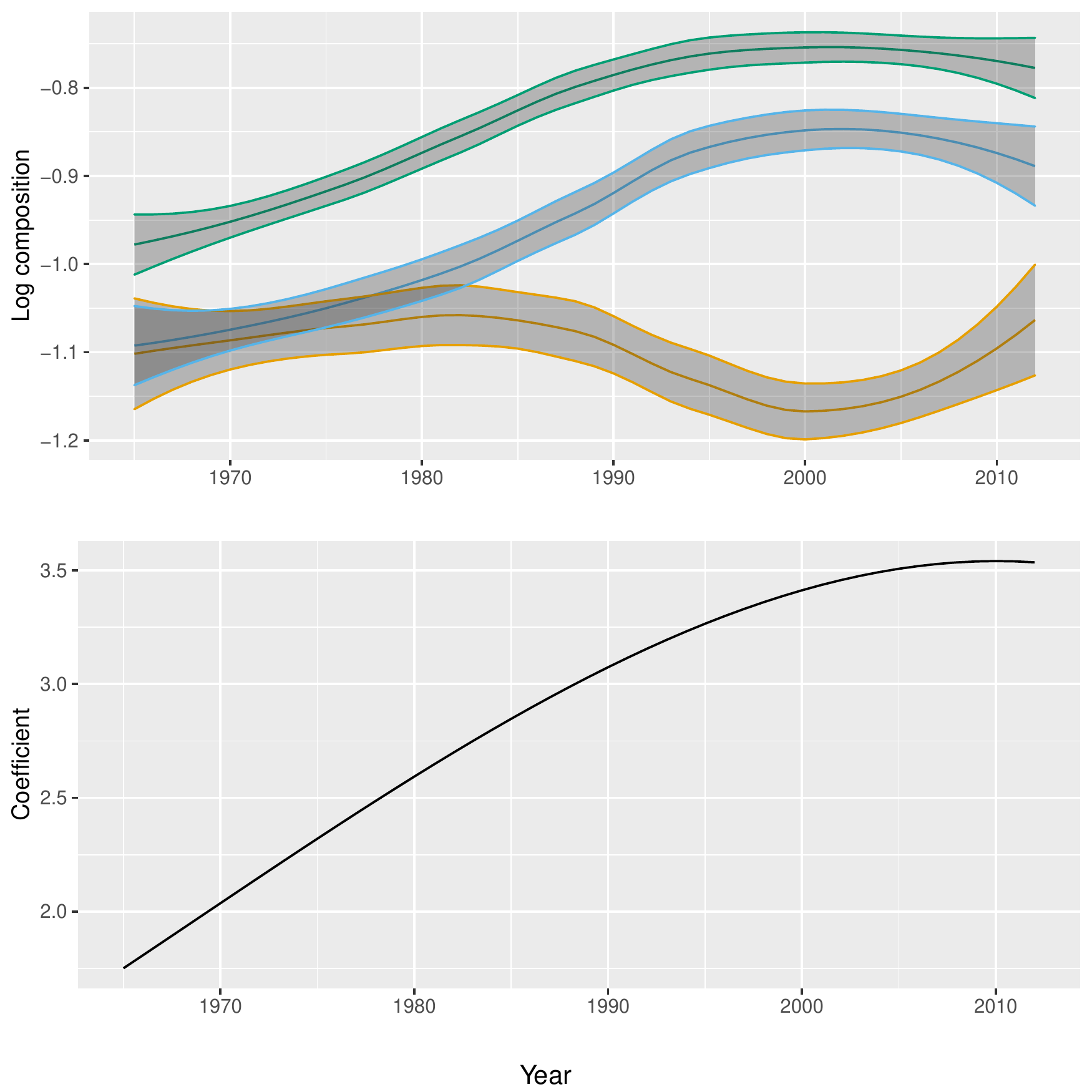}  
  \caption{NEOP, age class 40--64, females}
  \label{subfig:neop40-64F}
\end{subfigure}
\begin{subfigure}{0.45\textwidth}
  \centering
  \includegraphics[width=\linewidth]{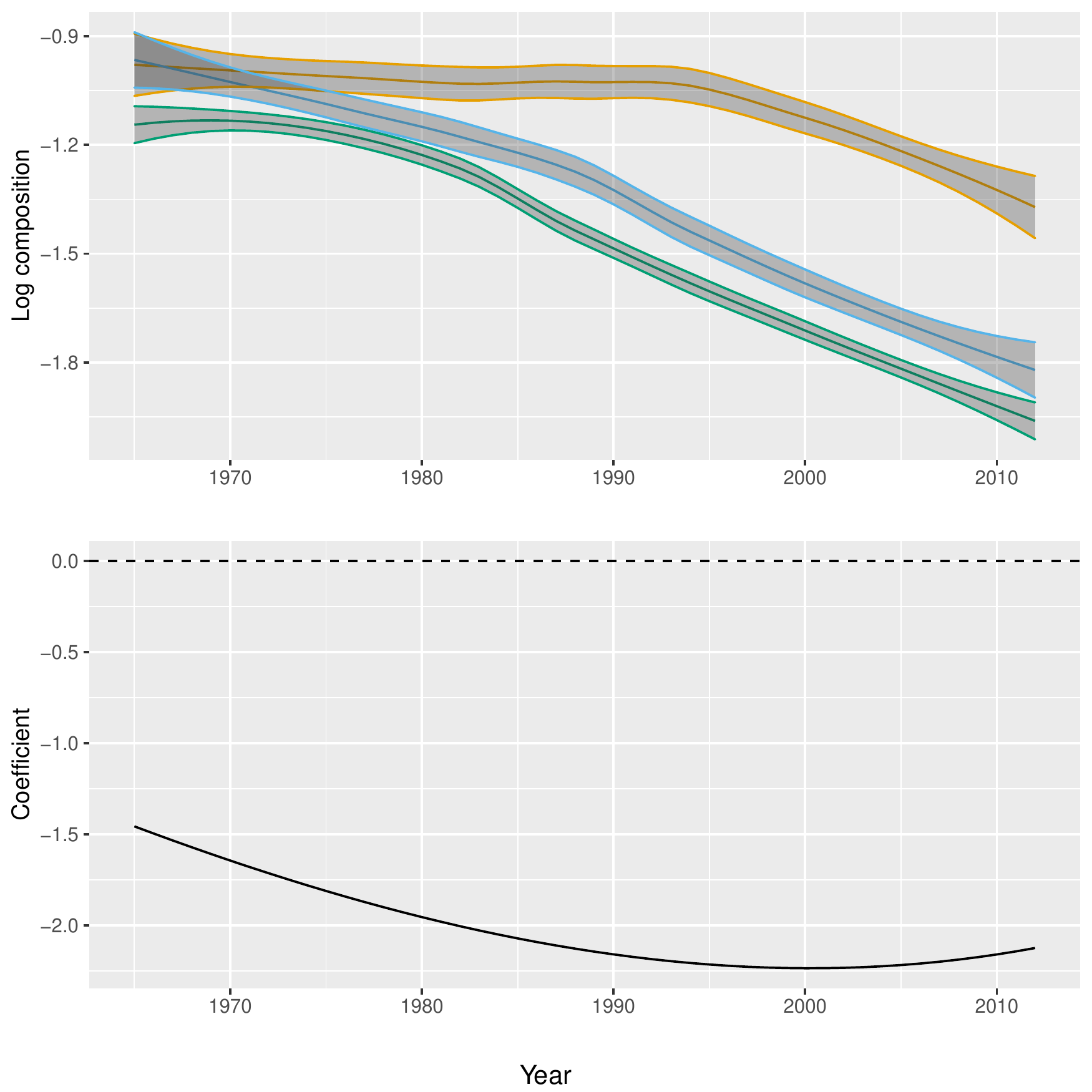}  
  \caption{CIRC, age class 40--64, females}
\end{subfigure}

\begin{subfigure}{0.45\textwidth}
  \centering
  \includegraphics[width=\linewidth]{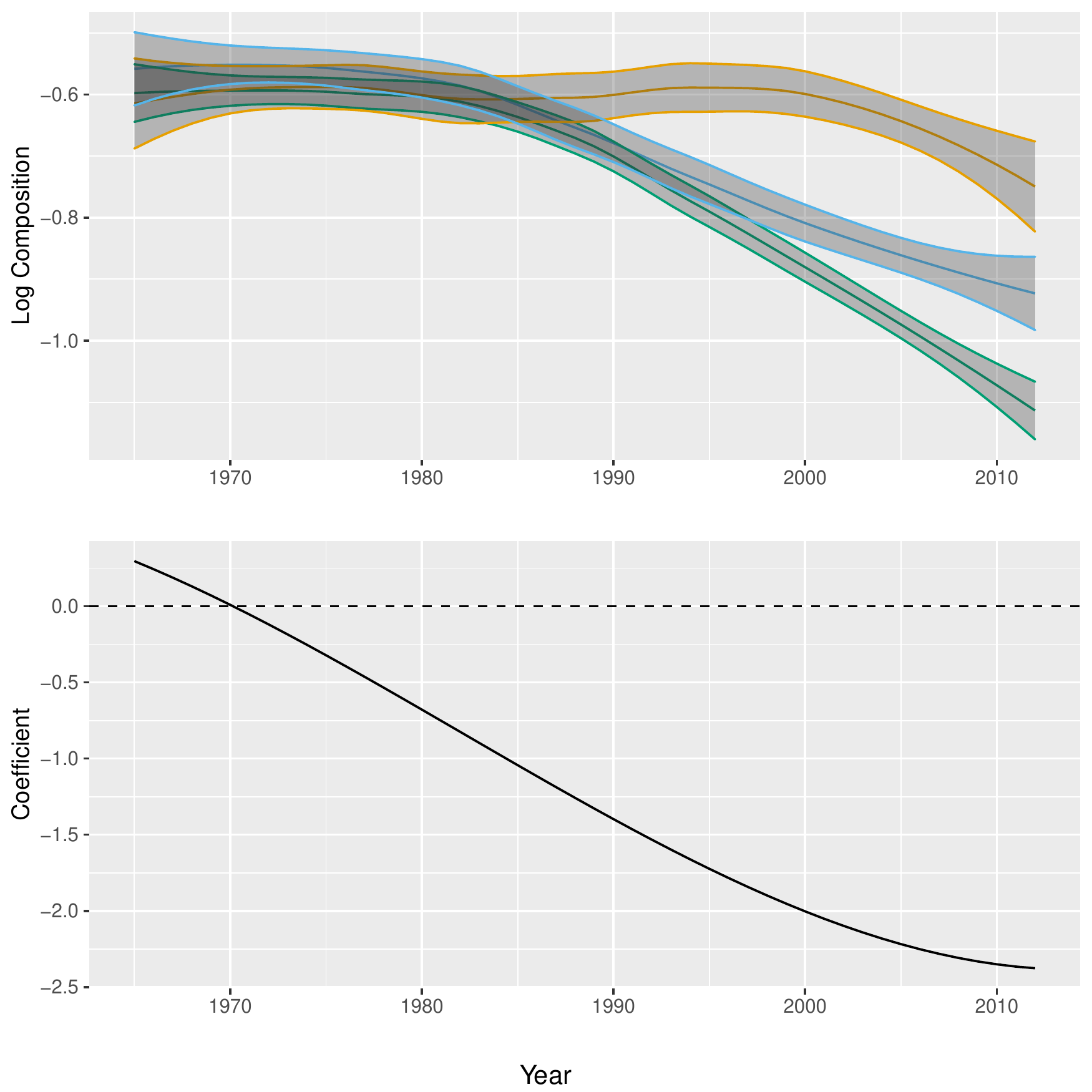}  
  \caption{CIRC, age class 65+, males}
\end{subfigure}
\begin{subfigure}{0.45\textwidth}
  \centering
  \includegraphics[width=\linewidth]{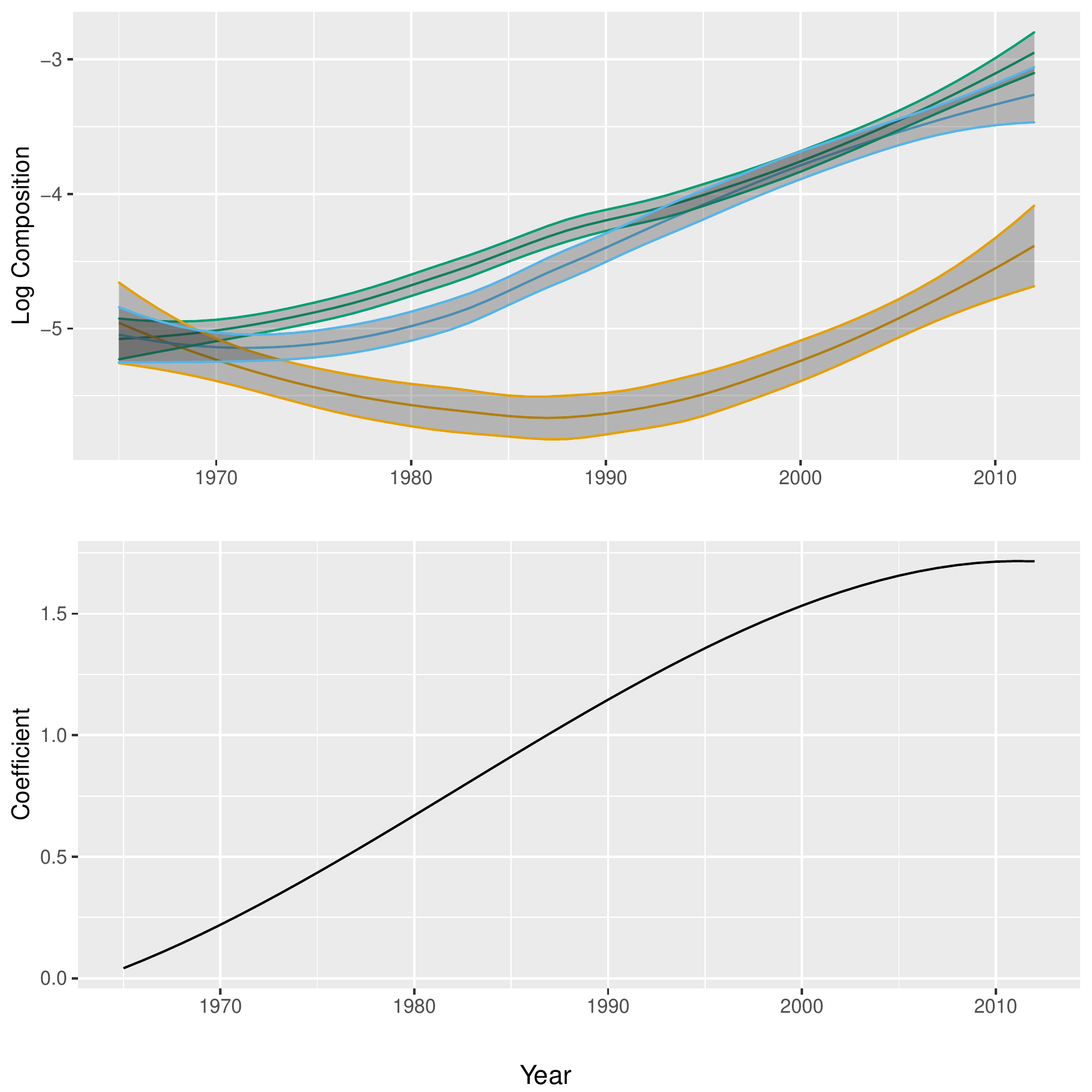}  
  \caption{NERV, age class 65+, males}
\end{subfigure}

\caption{\textit{Smoothed curves of log composition of some causes of death for three clusters of countries, with the estimated coefficient curves below. For each predictor and year, the nations are divided into three groups characterized by low (in yellow), medium (in light blue) and high (in green) life expectancy.}}
\label{fig:marg_eff}
\end{figure}

Regarding the results relative to specific causes, it is worth recalling that the interpretation of coefficients for the log-contrast model is different from the standard linear regression model. The main reason lies in the zero-sum constraint, which reflects the fact that one component increases its relative importance only if one or more of the others decreases \citep{coenders2020interpretations}. For the model (\ref{eq:model_func}), it can be shown that the following interpretation holds at each time $t$. Multiplying by a factor $c$ the ratio of one component $\beta_{jl}(t)$ of the $j$-th composition over each of the other parts $\beta_{jm}(t), m=1,\ldots,j-1,j+1,\ldots,p_j$ leads to a change of $\log(c)\beta_{jl}(t)$ in the expected value of the response variable. Equivalently, we can also interpret the coefficients jointly as follows. The expected value of the response variable grows when increasing the relative importance of components with positive coefficient and reducing that of components with negative coefficient. However, interpretation over time is not straightforward and we make use of additional plots to elucidate it, following \citet{sun2020}. The idea is to compare the smoothed trajectories of log compositions for three clusters of countries with the estimated coefficient curves. For each predictor and each year, the nations are divided into three groups characterized by low, medium and high life expectancy, thus giving rise to time-varying partitions. For each group, the smoothed values together with their $95\%$ confidence bands are calculated using local regression. In this way, we can also check whether our model describes relationships encountered in raw data. Figure \ref{fig:marg_eff} shows the resulting plots for four relevant causes. 
The graphs show that our model provides realistic results. We observe that increases (decreases) in the difference of the prevalence of a cause of death between high- and low-longevity countries are reflected in increasing (decreasing) coefficient curves. For example, considering the age class 40--64 for females, in the '60s, countries with higher prevalence of death by neoplasms and lower by circulatory diseases have higher life expectancy. In subsequent years, the difference in terms of prevalence of neoplasm between high- and low-longevity countries increases and this is reflected in the increasing estimated curve, while the reverse holds for circulatory diseases. 

\begin{figure}
    \centering
    \includegraphics[width=.9\textwidth]{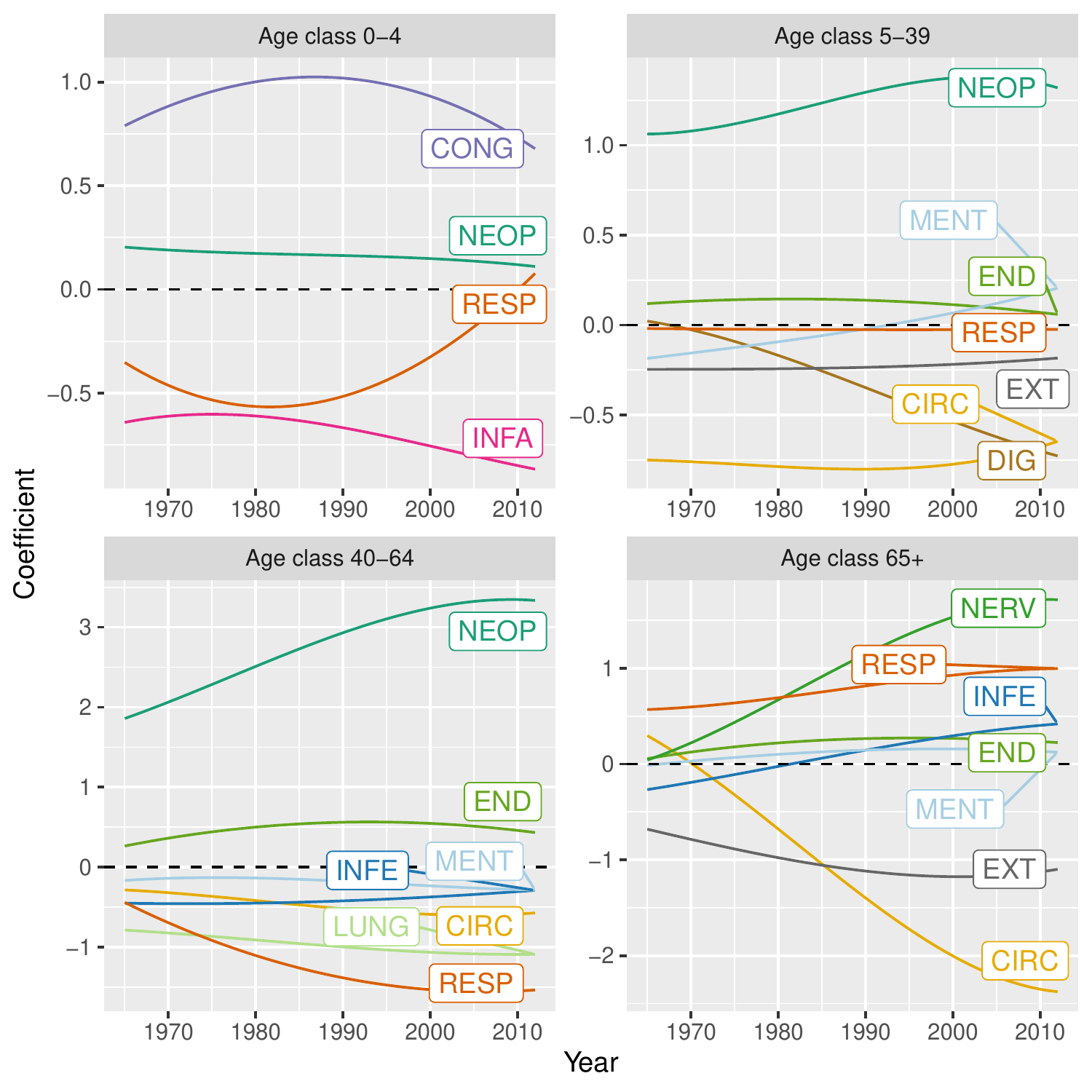}
    \caption{\textit{Estimated coefficient curves for the four age classes, males.}}
    \label{fig:coefM}
\end{figure}

The estimated coefficient curves for males are reported in Figure \ref{fig:coefM}. 
The positive increasing trend of neoplasms in age classes 5--39 and 40--64 is a clear effect due to substitute mortality, which has been defined as “that mortality which results from a decrease in another specific disease” \citep{van1997health}. That is, in many countries with high longevity, cancer mortality has become the main cause of death due to the reduction of other conditions, such as those related to the circulatory system. In fact, circulatory diseases can be seen to have a negative effect for all age groups, excluding 0--4. Another cause with a negative decreasing effect in age class 5--39 is digestive diseases. It can be linked to the high incidence of this class of diseases, particularly liver cirrhosis, observed in early adulthood for Eastern European nations \citep{blachier2013burden} and other Commonwealth countries, such as the UK \citep{lewer2020premature}. For the age class accounting for senescent mortality, the effect of circulatory diseases is negative and strongly increasing, concurrently with the positive increasing effect of nervous, respiratory and infectious diseases. These are conditions whose susceptibility is higher in the elderly. The estimated positive increasing effect reflects the process of population aging, that is, the increase in proportion of population aged 65 and over, which is particularly vulnerable to the aforementioned diseases. It is interesting to highlight the sign change of infectious diseases, which means that in the first period this condition was associated with low-longevity countries.

\begin{figure}
    \centering
    \includegraphics[width=.9\textwidth]{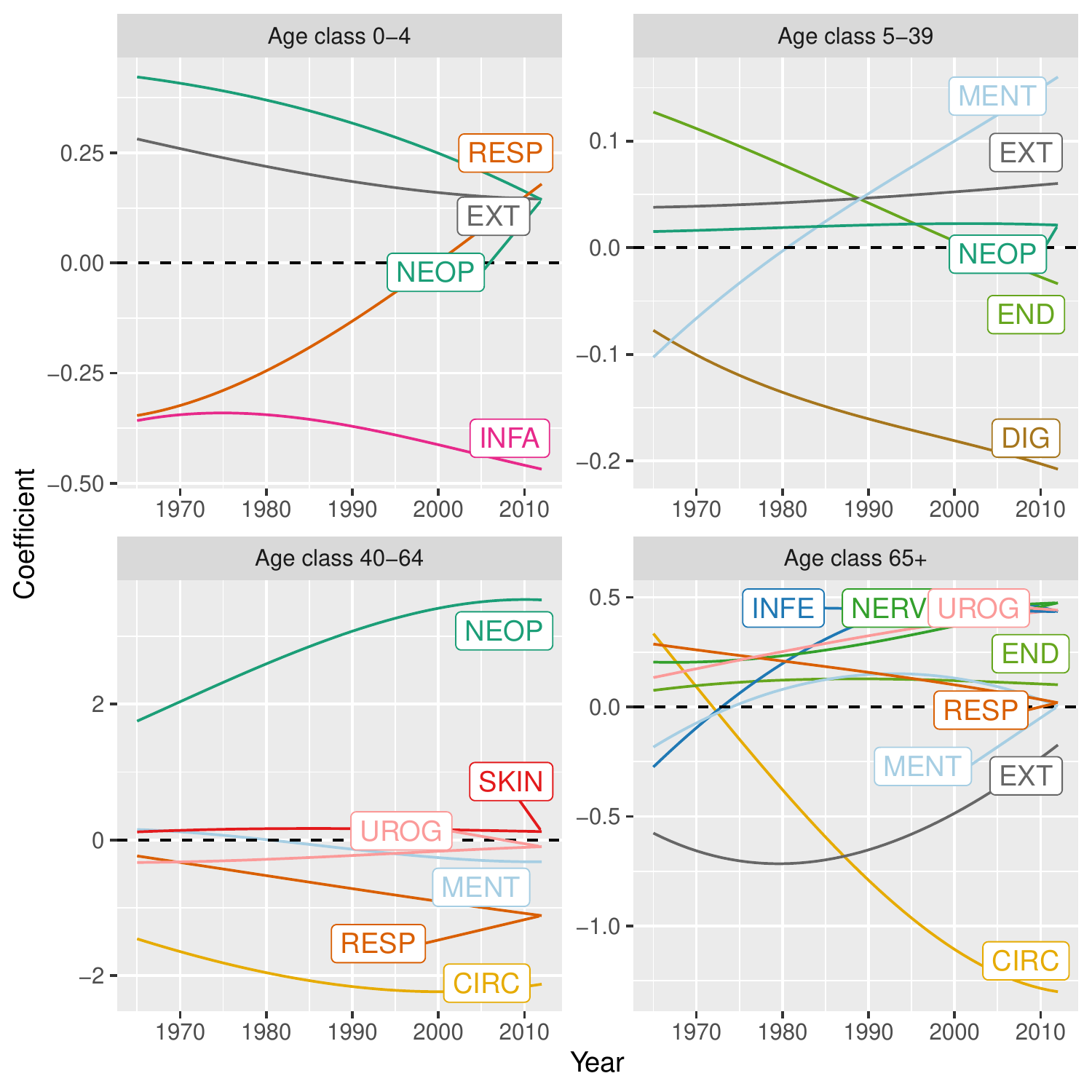}
    \caption{\textit{Estimated coefficient curves for the four age classes, females.}}
    \label{fig:coefF}
\end{figure}

The results for females are reported in Figure \ref{fig:coefF}. 
Compared to males, the increasing positive effect of neoplasms and the increasing negative effect of circulatory diseases in age class 40--64 overshadow all others in terms of magnitude. In this age group, differently to males, skin and urogenital diseases are selected. On the contrary, endocrine and infectious diseases, as well as lung cancer, are not included. One possible explanation for the non-inclusion of lung cancer is its high mortality rate in both low- and high-life expectancy countries for women \citep{jani2021lung}. In the senescent age group, the effect of respiratory diseases is positive decreasing and, unlike males, there is an increase in the prevalence of urogenital diseases over time for high-longevity countries. This cause, which is also selected for age classes 40-64, appears to be a sex-specific cause.
 
\begin{figure}
     \centering
     \includegraphics[scale = 0.7]{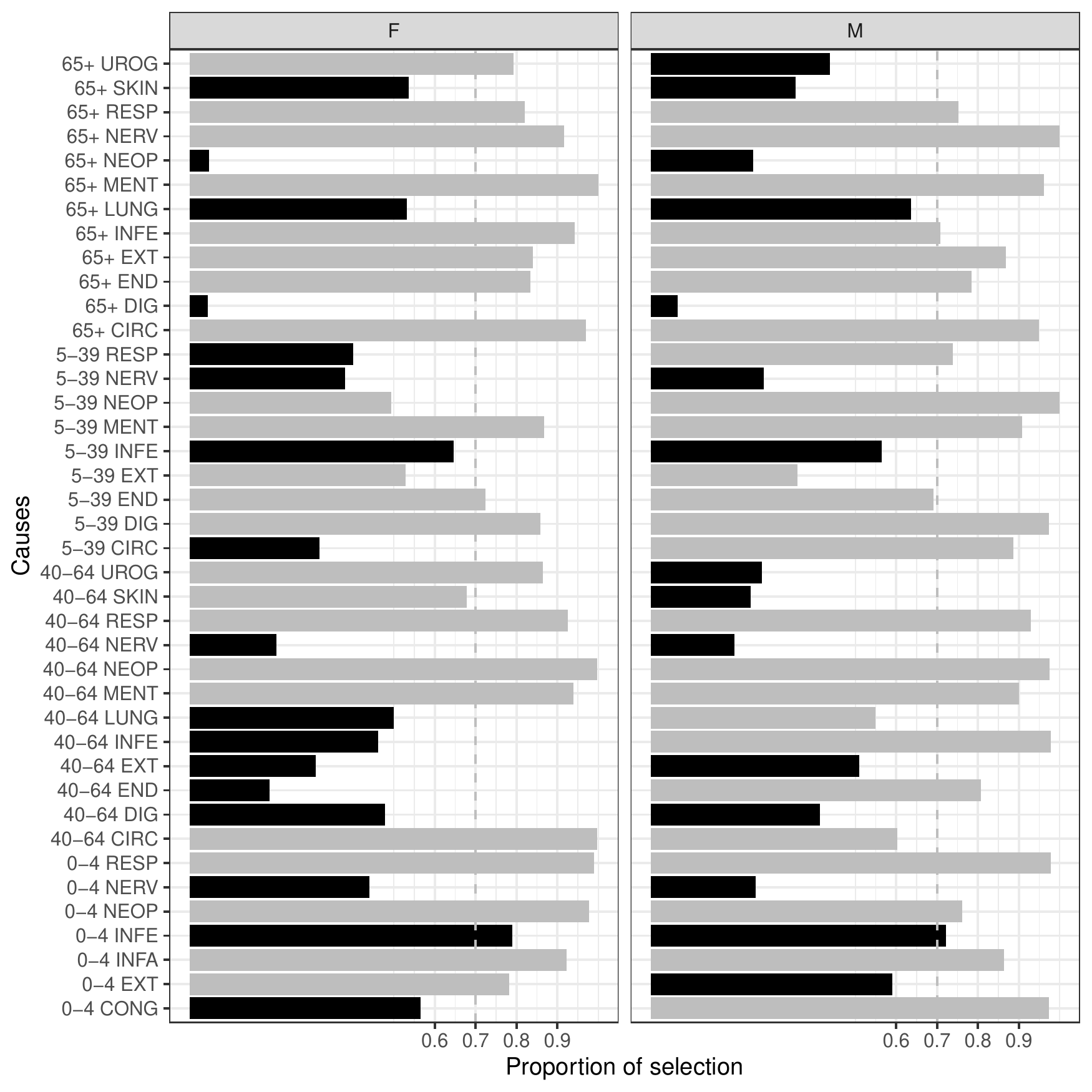}
     \caption{\textit{Proportion of the causes of death selected in 500 bootstrap samples, for females and males. In gray, the bars of the selected predictors from fitting the model to the original data, in black the bars of the estimated null coefficients.}}
     \label{fig:boot}
 \end{figure}

To assess the stability of the selection procedure, we generated 500 bootstrap samples and used leave-one-out cross-validation to select the tuning parameters, as for the model estimated with the original data. The results reported in Figure \ref{fig:boot} show that the variable selection is quite stable. In general, our proposal appears to be able to select the relevant predictors, at the cost of including some causes which may not have much effect on life expectancy. This is the case of external diseases in the age class 5--39 for both sexes, as well as neoplasms in the age class 5--39 for females and lung cancer and circulatory diseases in the age class 40--64 for males. On the other hand, infectious diseases for the age group 0--4 is selected in more than $70\%$ of the bootstrap samples for both sexes, indicating that it may play an important role, although its coefficient is estimated to be zero. 
 
\section{Discussion}\label{sec:disc}
We introduced a functional regression model with compositional covariates in the spirit of the proposal by \cite{sun2020}, extending their work to the relevant framework of a functional response. The model allows us to explain the evolution of life expectancy at birth for several countries as a function of the compositions derived from cause-specific mortality rates of four distinct age groups. The method involves a B-spline expansion of the unknown functional coefficients coupled with a group-Lasso penalty, enabling variable selection at the function level and consequently high interpretability of the results. The methodology is implemented within the \textsc{R} package \textsc{fcrc}, available at \url{https://github.com/emanuelegdepaoli/fcrc}, where the code for reproducing the analysis, the simulation studies and all images of the paper is also included.

It is worth noting that causes of death cannot be regarded as causal drivers of overall mortality (life expectancy). The main reason is that the cause of death and mortality occur simultaneously, so one cannot be the cause of the other. It would be more sensible to include risk factors (e.g. life–styles, pollution, etc.) to assess a causal link with mortality. However, to our knowledge, there is no harmonized and sufficiently high quality cross--country data over time on risk factors to do that. Instead, causes of death data are available, and while they cannot be considered really ``drivers'' of overall mortality we can see them as mediators between risk factors and life expectancy. Therefore an analysis such as ours can indirectly provide additional insights on the epidemiological trajectories of countries.

One major finding is that life expectancy is mainly driven by mortality at age 40--64 for women, while for men the 65+ and 5-39 age groups are also relevant. Not surprisingly, we found that circulatory diseases are increasingly relevant in determining the life expectancy of countries: the lower the relative importance of circulatory diseases, the higher the life expectancy. We also found an increasing relevance of digestive diseases for young men and women and of lung cancer for young men only. Other results, such as the increasingly positive effect of neoplasms at age 40--64 and of diseases of nervous system at age 65+ (that is, the higher the relative importance of these causes, the higher the life expectancy) can be explained in terms of ``substitution effect'', which means that the increasing relevance of these causes is an indirect effect of the reduction of other causes. We should keep in mind that the sample is made up of several countries with a different pattern of overall and cause-specific mortality. In particular, Eastern European countries that underwent a serious mortality crisis after the fall of the Soviet Union have a peculiar pattern that might have driven some of these results.

The proposed model allows us to simultaneously consider all causes of death and age groups in determining the evolution of overall mortality. This is increasingly important, since it has been observed that the composition of cause-specific mortality is becoming increasingly diversified \citep{Bergeron-Bouchere002414}, thus making analyses based on a single cause of death less reliable. 

We consider the summary measure of life expectancy at birth, but other measures such as the modal age at death \cite{canudas-romo2008}, which is not affected by infant mortality, or lifespan disparity \cite{canudas-romo-vaupel2003}, which is a measure of compression of age--specific mortality, can be used as a response variable, providing further insights on the evolution of mortality in high income countries.
%%%%%%%%%%%%%%%%%%%%%%%%%%%%%%%%%%%%%%%%%%%%%%
%% Single Appendix:                         %%
%%%%%%%%%%%%%%%%%%%%%%%%%%%%%%%%%%%%%%%%%%%%%%
%\begin{appendix}
%\section*{???}%% if no title is needed, leave empty \section*{}.
%\end{appendix}
%%%%%%%%%%%%%%%%%%%%%%%%%%%%%%%%%%%%%%%%%%%%%%
%% Multiple Appendixes:                     %%
%%%%%%%%%%%%%%%%%%%%%%%%%%%%%%%%%%%%%%%%%%%%%%
%\begin{appendix}
%\section{???}
%
%\section{???}
%
%\end{appendix}

%%%%%%%%%%%%%%%%%%%%%%%%%%%%%%%%%%%%%%%%%%%%%%
%% Support information, if any,             %%
%% should be provided in the                %%
%% Acknowledgements section.                %%
%%%%%%%%%%%%%%%%%%%%%%%%%%%%%%%%%%%%%%%%%%%%%%
\begin{acks}[Acknowledgments]
%The authors acknowledge financial support from the PRIN 2017 project ``Unfolding the SEcrets of LongEvity: Current Trends and future prospects'' (SELECT), project number 20177BRJXS
This research was supported by the PRIN 2017 project SELECT (20177BRJXS) and by the MUR-PRIN 2022 project CARONTE (2022KBTEBN), funded by the European Union - Next Generation EU. The authors also thank Emilio Zagheni, Ugofilippo Basellini and other scholars from the Max Planck Institute for Demographic Research for useful discussion during the presentation of earlier versions of this work.
% The authors would like to thank ...
\end{acks}
\bibliographystyle{imsart-nameyear} % Style BST file
\bibliography{biblio}       % Bibliography file (usually '*.bib')

%% or include bibliography directly:
% \begin{thebibliography}{}
% \bibitem[\protect\citeauthoryear{???}{???}]{b1}
% \end{thebibliography}

\end{document}